\documentclass[aps,tightenlines,longbibliography,notitlepage,nofootinbib]{revtex4-2}

\usepackage{amsmath}
\usepackage{amssymb}
\usepackage{bbm}
\usepackage{simplewick}
\usepackage{mathtools}
\usepackage[normalem]{ulem}
\usepackage{dsfont}
\usepackage{mathrsfs}
\usepackage[makeroom]{cancel}
\usepackage{graphicx}
\usepackage{bm}

\def\rhat{{\boldsymbol{\hat{r}}}}
\def\khat{{\boldsymbol{\hat{k}}}}

\DeclareMathOperator\erfc{erfc}

\DeclareMathOperator\BesselJ{J}
\DeclareMathOperator\SphBesselj{j}
\DeclareMathOperator\BesselK{K}
\DeclareMathOperator\StruveH{H}

\def\0{\boldsymbol{0}}
\def\RR{\mathbbm{R}}

\def\BD#1{{\boldsymbol{{#1}}}}

\def\ii{\mathrm{i}}
\def\bnabla{{\boldsymbol\nabla}}
\def\VAREPS#1{\varepsilon_{#1}}
\def\OM#1{\omega_{#1}}

\def\B#1{\left(#1\right)}
\def\BB#1{\left[#1\right]}

\def\la{\langle}
\def\ra{\rangle}
\def\lara#1{\la#1\ra}

\def\be{\begin{equation}}
\def\ee{\end{equation}}

\def\hc{\text{h.c.}}
\def\cc{\text{c.c.}}

\def\for{\ \text{for} \ }

\usepackage[ddmmyyyy]{datetime}
\makeatletter
\def\Dated@name{}
\makeatother

\def\ME{m\epsilon}

\def\pref{0.5}

\def\KS{{\BD{k}\sigma}}

\begin{document}
\title{Expanding quantum Proca field}
\author{Bogdan Damski}
\affiliation{\text{Jagiellonian University,
Faculty of Physics, Astronomy and Applied Computer Science,}
\\ {\L}ojasiewicza 11, 30-348 Krak\'ow, Poland}
\begin{abstract}
We consider the Proca theory of a real massive vector field coupled to an external current. 
This theory allows for non-conservation of the charge associated with the
external current. It is so  because the total conserved current
is given by the sum of the external current and the internal current
associated with the Proca  field, neither of which is required to be conserved separately.
 Moreover, this theory admits the 
quantum description of a longitudinal field generated by a charge distribution. 
We explore both features by studying the quantum dynamics underlying 
the causal formation of the Proca field following  
the charge-creation process.
The resulting non-equilibrium quantum state exhibits an expanding
Proca field accompanied by  a moving   front.
We compare this state with the one representing the corresponding 
static Proca field and identify fundamental differences between these 
states.
The quantization of the Proca theory in the presence of an external 
current is discussed in detail.
General results concerning the quantum dynamics of this theory are 
presented. Particular emphasis is placed on the problem of the instantaneous
switching-on of an otherwise static external current.
An intriguing shockwave structure is found in the point-charge creation problem,
the extended-charge creation problem is also discussed.
\end{abstract}
\maketitle

\def\jmat{\boldsymbol{j}_\text{mat}}
\def\jjmat{j_\text{mat}}

\section{Introduction}

We study the vector field theory defined by the following Lagrangian density
\be
{\cal L}=
-\frac{1}{4} F^{\mu\nu}F_{\mu\nu}
+\frac{m^2}{2}V^\mu V_\mu-\jjmat^\mu V_\mu,
\label{Lproca}
\ee
where $F^{\mu\nu}\equiv\partial^\mu V^\nu-\partial^\nu V^\mu$,
$V$ is the Proca  field, and $j_\text{mat}$ is the external current
(see Appendix \ref{Conventions_app}  for our conventions).
The particles described 
by this theory will be referred to as vector bosons
($m>0$ denotes their  mass).
This is the free Proca theory supplemented with the coupling of 
the Proca field to the external current. This theory 
appears in various  contexts, which we briefly explain 
below.

First, as is extensively discussed in review articles
\cite{Nieto_RMP1971,Gillies2005,Nieto_RMP2010}, the Proca theory 
is regarded as a viable extension of the Maxwell theory.
This view was recently challenged in \cite{NoProca}
but it remains to be seen whether the strong claim advanced 
there--following from the idealized choice of boundary conditions imposed on
a   Proca field propagating in a coaxial cable--is physically justified.
Second, basic properties of vector bosons,
such as  $\rho$ and $\omega$ mesons, are captured by the Proca theory 
\cite{Greiner,ColemanBook,Weinberg}.
Third, the Proca theory has a paradigmatic status in field theory,
where various concepts are illustrated using it
 \cite{Greiner,ColemanBook,Weinberg}.
Finally, 
we observe that
modified Proca theories are 
discussed in a cosmological context \cite{HeisenbergPhysRep2019}.

In the context of our work, 
theory (\ref{Lproca}) has the following remarkable features: 
(i) 
it enables a fully quantum description of the field associated with a given charge
and (ii) it admits non-conservation of the  charge.
The goal of this work is to explore these features.
For this purpose,  we will examine the quantum state representing 
the field of an omnipresent charge and the quantum state   encoding 
the field of a newly created charge (fundamental differences between these
states will be quantified). Before doing so, we briefly comment on
features (i) and (ii) as well as on  the relation of the present work to 
\cite{BDmag}.

Feature (i) follows from the fact that vector bosons
can have a longitudinal polarization in addition to the transverse ones.
Thereby, not only transverse but also longitudinal components of the  field
are quantized in  the Proca  theory. This is in stark contrast to the Coulomb-gauge Maxwell theory,
where the quantization procedure leaves classical the longitudinal component
of a field.

Feature (ii)  follows from field equations
\be
\partial_\mu  F^{\mu\nu}=\jjmat^\nu-m^2 V^\nu,
\label{ELproc}
\ee
which upon the action of $\partial_\nu$ yield 
\be
\label{total}
\partial\cdot(\jjmat - m^2V)=0.
\ee
This shows
that there is the total  conserved    current in the discussed 
theory, which is given by
the sum of the external current $\jjmat$ and the internal current $-m^2V$.
As a result of that, neither the external nor internal current has to be separately conserved,
which opens the way to considering the charge creation process.
This is markedly different from
the situation in the Maxwell theory,
where the contribution   $-m^2V$  is non-existent and so $\partial\cdot\jjmat$
must vanish.\footnote{One may 
still formally consider charge creation within the classical Maxwell theory, which is discussed in 
\cite{BermanAmJPhys2008}.
In this questionable approach, it is assumed that at some instant a conserved  external current 
appears, which transfers an electric charge all the way from spatial infinity to a given point in space.}
We mention in passing that charge conservation in the Maxwell theory is rooted  in relativistic
invariance and photon masslessness  (see e.g. \cite{Weinberg1964}).

Regarding charge non-conservation, suitable modifications of the Standard Model have been discussed
in the literature (see e.g. review articles \cite{Okun1989,Nieto_RMP2010}).
The aim of these works was to describe processes such as the decay of an 
electron into neutral particles (e.g. neutrinos, photons). In contrast to these works,
we adopt the 
external-current approach, where  the mechanism responsible for charge non-conservation 
is left unspecified. Using this approach, we investigate the dynamics of the 
resulting field configuration,
which to the best of our knowledge has not been addressed previously.
In doing so, we focus our attention on a Proca field
itself rather than on the related
``electromagnetic'' field, which appears below merely  in 
the discussion of the  quantization procedure.

Finally, we note that we recently studied the dynamics of the electromagnetic
field following the sudden switching-on of a  conserved external current
in the  quantum Coulomb-gauge Maxwell  theory \cite{BDmag}. This system differs from 
the one studied in the present work in virtually every physical aspect.
Still, we find some general features of the two systems to be similar, if not identical. 
On the one hand, this points to a certain universality of the quantum dynamics 
following the sudden onset of a static external current, which we find remarkable. 
On the other hand, this motivates the organization of  the discussion in the present work 
in a manner similar to that of \cite{BDmag}, which should facilitate
understanding  similarities and differences  in the    two problems.
In response to the latter observation, 
Secs.~\ref{Static_sec} and \ref{Problem_sec}, presenting  results derived
without assuming the specific form of an  external current,
are organized similarly to the corresponding sections of \cite{BDmag}. 
The ground state of the theory in the presence of a static external
current is constructed in Sec.~\ref{Static_sec}. 
The non-equilibrium state resulting from the sudden switching-on 
of an otherwise static external current is discussed in
Sec.~\ref{Problem_sec}.
Secs.~\ref{Quantization_sec}, \ref{Point_sec}, and \ref{Extended_sec} either have no 
counterpart in \cite{BDmag} or are necessarily organized differently from the 
corresponding discussion therein. The quantization of the Proca theory in 
the presence of an external current is discussed in Sec.~\ref{Quantization_sec}. 
The dynamics of the Proca field resulting from 
point-charge creation is discussed in Sec.~\ref{Point_sec}.
This analysis is broadened to the case of  
 extended-charge creation in Sec.~\ref{Extended_sec}. The 
summary of our work is presented in Sec.~\ref{Summary_sec}.

\section{Quantization}
\label{Quantization_sec}

A particularly comprehensive discussion of the 
quantization of the Proca theory  in the absence of an
external current can be found in \cite{Greiner}.  
The presence of a non-conserved  external current   requires  
a careful re-examination of the quantization procedure (e.g. because
the $4$-divergence of the Proca field operator no longer vanishes).
Therefore, we proceed step by step rather than jumping 
directly to the final result. Before doing so,  we introduce 
$\BD{E}=-\partial_0\BD{V}-\bnabla V^0$
and $\BD{B}=\bnabla\times\BD{V}$. 
In terms of the vector field, 
these quantities are defined in the same way as the electric and magnetic 
field in the Maxwell theory. Therefore, even though we are  working
with a massive rather than a massless theory, we will use the same notation.

To begin, $\partial{\cal L}/\partial(\partial_0 V^i)$,  the momentum canonically
conjugate to $V^i$,  
equals $-E^i$. Thereby, 
\begin{align}
\label{VEcom}
&[V^i(t,\BD{r}),E^j(t,\BD{r}')]=-\ii \delta_{ij}\delta(\BD{r}-\BD{r}'),\\
& [V^i(t,\BD{r}),V^j(t,\BD{r}')]=
[E^i(t,\BD{r}),E^j(t,\BD{r}')]=0.
\label{VVEEcom}
\end{align}
Then, we stress  that field equations (\ref{ELproc}) imply 
a field constraint
\be
V^0(t,\BD{r};t)=-\frac{\bnabla\cdot\BD{E}(t,\BD{r})  }{m^2} +
\frac{\jjmat^0(t,\BD{r})}{m^2},
\label{V0j}
\ee
where the first (third) argument of $V^0$ refers to the time dependence originating
from the quantum operator (the classical external current). We will use such double time notation for
operators that are defined as functions of both the external current $\jjmat$ and the canonical 
variables $\BD{V}$ and $-\BD{E}$. 
For these operators, a distinction between the time argument of operators and that of the external current
is necessary  when one works with Schr\"odinger-picture 
expressions. Specifically, the 
Schr\"odinger-picture version of $V^0$ is $V^0(0,\BD{r};t)$ 
because the unitary transformation from the Heisenberg to the 
Schr\"odinger picture leaves $\jjmat^0(t,\BD{r})$ unchanged.
Two related comments are in order. First, $V^0(t,\BD{r};t)$, 
$\BD{V}(t,\BD{r})$, $\BD{E}(t,\BD{r})$, and $\BD{B}(t,\BD{r})$
are Heisenberg-picture operators. Second, from the perspective
of the canonical formalism, $\bnabla\cdot\BD{E}$ and $\BD{B}$ are functions
of the canonical variables. Therefore, like the canonical variables themselves, 
they carry  a single time argument.

Moving on, a standard computation leads to the following 
Schr\"odinger-picture Hamiltonian expressed in terms of 
the canonical variables and the external current\footnote{One encounters a similar situation
while describing the interaction of a quantum  system
with a classical electromagnetic wave, where the Schr\"odinger-picture
Hamiltonian is explicitly time dependent.
}
\begin{align}
&H(0;t)= H_0+H_1(0;t),\\
\label{HH00}
&H_0= \frac{1}{2}\int d^3r\B{\BD{E}(0,\BD{r})\cdot\BD{E}(0,\BD{r}) +
\BD{B}(0,\BD{r})\cdot\BD{B}(0,\BD{r}) + m^2 \BD{V}(0,\BD{r})\cdot\BD{V}(0,\BD{r})
+(\bnabla\cdot\BD{E}(0,\BD{r})/m)^2},\\
&H_1(0;t)= \int d^3r \B{-\frac{ \jjmat^0(t,\BD{r})  }{m^2}\bnabla\cdot\BD{E}(0,\BD{r}) -
\BD{\jjmat}(t,\BD{r})\cdot\BD{V}(0,\BD{r})} + \zeta(t),
\label{H10t}
\end{align}
where 
\be
\zeta(t)=\frac{1}{2}\int d^3r  \B{\frac{\jjmat^0(t,\BD{r})}{m}}^2
\ee
and 
the first (second) argument of $H$ and $H_1$ is introduced just as the 
 first (third) argument of 
 $V^0$.
We observe that $\zeta(t)$ can be interpreted as the energy
arising from contact interactions between the charges described by 
the external current, the interactions whose strength scales 
as $1/m^2$.

The above splitting of the Hamiltonian into free-field and interaction 
parts, $H_0$ and $H_1$, leads to the following definition 
of  interaction-picture operators 
\begin{align}
\label{nklde}
&H^I_1(t;t)=\exp(\ii H_0t)H_1(0;t) \exp(-\ii H_0t),\\
&V^0_I(t,\BD{r};t)= \exp(\ii H_0t)V^0(0,\BD{r};t)\exp(-\ii H_0t),\\
&O_I(t,\BD{r})= \exp(\ii H_0t) O(0,\BD{r})\exp(-\ii H_0t) \for O=\BD{V},
\BD{E}, \BD{B},
\end{align}
where the label $I$ denotes interaction-picture quantities.
Using these expressions and the canonical commutation relations, we arrive at 
\begin{align}
\label{d0EI}
& \partial_t\BD{E}_I(t,\BD{r})= \bnabla\B{\bnabla\cdot\BD{V}_I(t,\BD{r}) }
-\Delta\BD{V}_I(t,\BD{r}) + m^2 \BD{V}_I(t,\BD{r}),\\
& V^0_I(t,\BD{r};t)= -\frac{\bnabla\cdot\BD{E}_I(t,\BD{r})  }{m^2} +
\frac{\jjmat^0(t,\BD{r})}{m^2},
\label{VI000}
\end{align}
where the last equation still involves the contribution 
from the external current.
To map the interaction-picture setting  to the one corresponding to the free  theory,
we introduce 
\be
\label{Vfree}
V_\text{free}(t,\BD{r})\equiv\B{-\bnabla\cdot\BD{E}_I(t,\BD{r})/m^2,\BD{V}_I(t,\BD{r})}.
\ee
In terms of this quantity, we  obtain 
\begin{align}
\label{H1tt}
&H^I_1(t;t)= \int d^3r \jjmat(t,\BD{r}) \cdot V_\text{free}(t,\BD{r}) + \zeta(t),\\
\label{V000}
& V^0_I(t,\BD{r};t) = V^0_\text{free}(t,\BD{r})  + \frac{\jjmat^0(t,\BD{r})}{m^2}, \\ 
\label{VIII}
& \BD{V}_I(t,\BD{r}) = \BD{V}_\text{free}(t,\BD{r}),\\
\label{BIII}
&\BD{B}_I(t,\BD{r})= \bnabla\times \BD{V}_\text{free}(t,\BD{r}),\\
& \BD{E}_I(t,\BD{r})=-\partial_t \BD{V}_\text{free}(t,\BD{r})
-\bnabla V^0_\text{free}(t,\BD{r}),
\label{EIII}
\end{align}
where (\ref{H1tt})--(\ref{BIII}) trivially follow from (\ref{Vfree})
while (\ref{EIII}) has been obtained in the following way.
We introduce the interaction-picture evolution operator
\be
U_I(t,0) = \mathbbm{T}\exp\B{-\ii \int_0^t dt' H_1^I(t';t')},
\label{UI}
\ee
where $\mathbbm{T}$ is the time-ordering operator. Next, we observe that
$ \BD{E}_I(t,\BD{r})=U_I(t,0) \BD{E}(t,\BD{r}) U^\dag_I(t,0)$,
where we substitute 
$ \BD{E}(t,\BD{r}) = -\partial_t (U^\dag_I(t,0) \BD{V}_I(t,\BD{r}) U_I(t,0))
-  U^\dag_I(t,0) \bnabla V^0_I(t,\BD{r};t) U_I(t,0)$. Standard 
manipulations involving the canonical commutation relations
then lead to  (\ref{EIII}).

Putting (\ref{VIII}) and (\ref{EIII}) into (\ref{d0EI})
as well as (\ref{EIII})  into (\ref{Vfree}), 
we find that  
\be
\partial_\mu  (\partial^\mu V^\nu_\text{free}-
\partial^\nu V^\mu_\text{free})=-m^2 V^\nu_\text{free},
\label{ELfree}
\ee
which is just  as in the  free Proca theory.
The same can be said about the equal-time commutation relations 
involving $\BD{V}_\text{free}$ and 
 $-\partial_0 \BD{V}_\text{free}-\bnabla V^0_\text{free}$,
 which is seen from the interaction-picture version of
the canonical commutation relations.
 Thereby, the textbook 
expression provides the representation of  $V_\text{free}$.
Namely, following \cite{Greiner}, we write it 
as\footnote{The vector field operator 
of the free Proca theory necessarily satisfies the Lorenz-gauge condition
($\partial\cdot V_\text{free}=0$). 
For completeness, we note that the gauge-invariant 
version of the free Proca theory  exists
(see e.g. \cite{Guralnik1968,Pimentel2015,BDGaugeinv1}).
}
\begin{align}
\label{A0r}
&V_\text{free}(t,\BD{r})= \int
\frac{d^3k}{(2\pi)^{3/2}}
\frac{1}{\sqrt{2\VAREPS{k}}}
\sum_\sigma \eta(\BD{k},\sigma)a_\KS\exp(-\ii\VAREPS{k}t+\ii\BD{k}\cdot\BD{r})
+\text{h.c.},\\
& \lbrack a_{\BD{k}\sigma},a^\dag_{\BD{k}'\sigma'}\rbrack=\delta_{\sigma\sigma'}\delta(\BD{k}-\BD{k}'), 
 \ \lbrack a_{\BD{k}\sigma},a_{\BD{k}'\sigma'}\rbrack=0,\\
 &\sigma,\sigma'=1,2:  \eta(\BD{k},\sigma)=\B{0,\BD{\eta}(\BD{k},\sigma)}, \ \BD{\eta}(\BD{k},\sigma)\in\RR^3, 
\ \BD{\eta}(\BD{k},\sigma)\cdot\BD{k}=0, 
\ \BD{\eta}(\BD{k},\sigma)\cdot\BD{\eta}(\BD{k},\sigma')=\delta_{\sigma\sigma'},\\
&\eta(\BD{k},3)=\B{\frac{\OM{k}}{ m},\frac{\VAREPS{k}}{ m}\khat},
\label{k3}
\end{align}
where $\VAREPS{k}=\sqrt{m^2+\OM{k}^2}$,  $\OM{k}=|\BD{k}|$, 
$\sum_\sigma\equiv\sum_{\sigma=1}^3$,
$\sigma=3$ ($\sigma=1,2$)
refers to the longitudinal (transverse) vector boson  polarization(s),
and all $a_\KS$ operators annihilate the vacuum state $|0\ra$.
This  state is the ground state of the free-field Hamiltonian, 
which in this  representation reads
\be
H_0= \int d^3k \VAREPS{k}\sum_{\sigma} a^\dag_\KS a_\KS.
\ee
This expression is obtained by (i) substituting (\ref{A0r}) into
(\ref{VIII})--(\ref{EIII}); (ii) setting $t=0$ in the resulting expressions
to obtain $\BD{V}(0,\BD{r})$, $\BD{B}(0,\BD{r})$, and $\BD{E}(0,\BD{r})$,
which are subsequently inserted into (\ref{HH00});  (iii) imposing normal
ordering.

\section{Time-independent  external current}
\label{Static_sec}

We discuss below the ground state of the system in the presence 
of a time-independent external current $\jjmat(\BD{r})$. 
For this purpose, we adopt the standard formalism employed
in \cite{BDmag}, which leads to the expected conclusion that
the ground state is a coherent state. Coherent states are frequently 
discussed in the context of quantum optics \cite{Gerry,Scully}, where
one deals with massless rather than massive vector bosons.
Nevertheless, their treatment in the present case proceeds
by the same token.

In the static setting, we deal with the following normal-ordered
Schr\"odinger-picture Hamiltonian 
\begin{align}
\label{Hdiag}
& H=\sum_\KS \VAREPS{k}  (b_\KS-\lambda_\KS)^\dag (b_\KS-\lambda_\KS)
-\sum_\KS \VAREPS{k} |\lambda_\KS|^2 + \zeta,\\
& [b_\KS,b^\dag_{\BD{k}'\sigma'}]= \delta_{\sigma\sigma'}  \delta_{\BD{k}\BD{k}'},
\  [b_\KS,b_{\BD{k}'\sigma'}]=0,
\end{align}
where we have moved from the continuum quantization model  to the periodic-box  
one through standard relations (see Appendix~\ref{Conventions_app}  for
 their summary).  Moreover, 
\begin{align}
\label{zetA}
&\zeta=  \frac{1}{2} \int d^3r\B{\frac{\jjmat^0(\BD{r})}{m}}^2,\\
&\lambda_\KS=-\frac{J_\text{mat}(\BD{k})\cdot\eta(\BD{k},\sigma)}{\sqrt{2V}\VAREPS{k}^{3/2}},
\label{lLKS}\\
\label{Jdef}
&J_\text{mat}(\BD{k}) =  \int d^3r \jjmat(\BD{r})\exp(-\ii\BD{k}\cdot\BD{r}),
\end{align}
where $V$ is the volume of space in the quantization-in-a-periodic-box scheme.
The absence of arguments in $H$ and $\zeta$ above, as well as in
$H_1$ below, reflects the static nature of the problem  discussed 
in this section.

The ground state of $H$, $|\tilde{0}\ra$ annihilated by all operators $b_\KS-\lambda_\KS$,
is given by  
\begin{align}
\label{lam}
&|\tilde{0}\ra= \bigotimes_\KS |\lambda_\KS\ra, \\
&b_\KS |\lambda_\KS\ra = \lambda_\KS |\lambda_\KS\ra,\\
&|\lambda_\KS\ra= D(\lambda_\KS) |0\ra,\\
&D(\lambda_\KS)=\exp(\lambda_\KS b^\dag_\KS - \lambda^*_\KS b_\KS),
\end{align}
and so it is a coherent state (see Appendix \ref{Conventions_app} for
the  summary of properties of coherent states).
The Schr\"odinger-picture wavefunction, energy, and its standard deviation,
all associated with the ground state,  
are\footnote{The standard deviation of a quantity, represented by 
an operator  $O$, is defined 
as $\sqrt{\lara{O^2}-\lara{O}^2}$, where $\lara{\cdots}$ denotes 
the expectation value in a discussed state.}
\begin{align}
 \label{psiTt}
&|\tilde{\psi}(t)\ra=\exp(-\ii\tilde{\cal E} t)|\tilde{0}\ra,\\
\label{tildeE}
&\tilde{\cal E}=-\sum_\KS \VAREPS{k}  |\lambda_\KS|^2+\zeta,\\
&\tilde{\sigma}_{\cal E}=0,
\end{align}
where the sum over polarizations  can be carried out via the completeness 
relation \cite{Greiner}
\be
\sum_\sigma \eta^\mu(\BD{k},\sigma)\eta^\nu(\BD{k},\sigma)
= -\eta^{\mu\nu}+\frac{(\VAREPS{k},\BD{k})^\mu (\VAREPS{k},\BD{k})^\nu}{m^2}.
\ee
We shall not use this relation 
because    (i) for a general external current discussed in  Secs.~\ref{Static_sec} and
\ref{Problem_sec} one deals with more compact expressions  without
performing the polarization sums; (ii) 
 for the specific external currents considered in
Secs.~\ref{Point_sec} and \ref{Extended_sec} these sums are trivial.

Then, we  observe that 
\begin{align}
\label{h0h1a}
&\lara{\tilde{\psi}(t)|H_0|\tilde{\psi}(t)}= \sum_\KS \VAREPS{k}  |\lambda_\KS|^2,\\
&\lara{\tilde{\psi}(t)|H_1-\zeta|\tilde{\psi}(t)}= -2 \sum_\KS \VAREPS{k}  |\lambda_\KS|^2,
\label{h0h1b}
\end{align}
which can obtained either from the above expressions or 
via the so-called Hellmann–Feynman theorem, which is discussed in
Appendix \ref{Hellmann_app}. This implies that  
\be
\lara{\tilde{\psi}(t)|H_1-\zeta|\tilde{\psi}(t)}= -2
\lara{\tilde{\psi}(t)|H_0|\tilde{\psi}(t)},
\label{H0H1B}
\ee
which has the virial theorem-like structure. 
Note, however, that (\ref{H0H1B}) is derived for the ground state and 
does not hold for excited eigenstates, which can be verified by 
adopting  the discussion presented in \cite{BDmag}. This contrasts with 
the virial theorem relation for the hydrogen atom, where  the expectation value
of the electron potential energy equals $-2$ times the expectation value of the electron kinetic energy 
in every eigenstate of the hydrogen atom Hamiltonian.

Finally, we introduce the vector-boson number operator and
give the corresponding expressions for the number of vector
bosons and its standard deviation in the state $|\tilde{\psi}(t)\ra$
\begin{align}
&N=\sum_\KS b_\KS^\dag b_\KS,\\
\label{phn}
&\tilde{\cal N}=\lara{\tilde{\psi}(t)|N|\tilde{\psi}(t)}
=\sum_\KS |\lambda_\KS|^2,\\
&\tilde{\sigma}_{\cal N}=\sqrt{\tilde{\cal N}}.
\label{VarN}
\end{align}
Moreover, the static  Proca field, $\lara{\tilde{\psi}(t)|V^0(0,\BD{r})|\tilde{\psi}(t)}=\jjmat^0(\BD{r})/m^2+
\lara{\tilde{\psi}(t)|V^0_\text{free}(0,\BD{r})|\tilde{\psi}(t)}$ and 
$\lara{\tilde{\psi}(t)|\BD{V}(0,\BD{r})|\tilde{\psi}(t)}=
\lara{\tilde{\psi}(t)|\BD{V}_\text{free}(0,\BD{r})|\tilde{\psi}(t)}$, reads
\begin{align}
\label{Vstat0}
&\lara{\tilde{\psi}(t)|V^0(0,\BD{r})|\tilde{\psi}(t)}
= \frac{\jjmat^0(\BD{r})}{m^2}+  
\sum_{\BD{k}} 
\frac{\OM{k}}{\sqrt{2V\VAREPS{k}}m}\B{\lambda_{\BD{k}3}\exp(\ii\BD{k}\cdot\BD{r})+\cc},\\
&\lara{\tilde{\psi}(t)|\BD{V}(0,\BD{r})|\tilde{\psi}(t)}=
\sum_\KS
\frac{\BD{\eta}(\BD{k},\sigma) }{\sqrt{2V\VAREPS{k}}}\lambda_\KS \exp(\ii\BD{k}\cdot\BD{r})+\cc
\label{VstatVec}
\end{align}

\section{Problem setup}
\label{Problem_sec}
We are now  interested in the situation, where
\be
\jjmat(t,\BD{r})=\jjmat(\BD{r})\theta(t)
\label{ewfkmlk}
\ee
and the system is initially, i.e. at $t=0$,  in the vacuum state $|0\ra$;
$\theta(t)$ is  the Heaviside step function.
To proceed, we  use 
the results of Appendix \ref{Time_app},
where the dynamics induced by a general time-dependent 
external current is analyzed. 
This yields the following coherent state-based 
Schr\"odinger-picture wavefunction
  for $t\ge0$
\begin{align}
\label{weqws}
&|\psi(t)\ra
=\exp(-\ii \zeta t)\bigotimes_\KS \exp\B{\ii\text{Im}\Phi_\KS(t)}
|\alpha_\KS(t)\exp(-\ii\VAREPS{k}t) \ra,\\
\label{alfaG}
&\alpha_\KS(t)= \B{\exp(\ii\VAREPS{k}t)-1}\lambda_\KS,\\
&\Phi_\KS(t)=\B{1+\ii \VAREPS{k}t -\exp(\ii \VAREPS{k}t)}|\lambda_\KS|^2.
\label{PhiG}
\end{align}
This wavefunction describes an expanding field, which is intuitively plausible
for causal reasons and will be illustrated through specific examples
in Secs.~\ref{Point_sec} and \ref{Extended_sec}.
In fact, state (\ref{weqws}) is the expanding  counterpart of ``static'' state (\ref{psiTt}).
Thereby, it is  unsurprising that $\lambda_\KS$ appears in the above expressions.
In the following, we compare these states   and discuss several 
quantities  that can be extracted out of  $|\psi(t)\ra$. 
While this analysis is carried out without assuming any particular form 
of $\jjmat(\BD{r})$, it implicitly assumes that the expressions under consideration are 
finite.

Finally, we note that we will use below the following Riemann–Lebesgue lemma-based
result
\be
\lim_{t\to\infty} \int_0^\infty  d\OM{k} f(\OM{k})\exp(\pm\ii\VAREPS{k}t)=0
\label{RLL}
\ee
 for real-valued $f\in L^1(0,\infty)$; see Appendix \ref{RL_app} for its
discussion.

\subsection{Proca field}
\label{Vector_sub}
Simple calculations lead to  
\begin{align}
\label{v000}
&\lara{\psi(t)|V^0(0,\BD{r};t)|\psi(t)}=
\lara{\tilde{\psi}(t)|V^0(0,\BD{r})|\tilde{\psi}(t)}
-\sum_\BD{k} \frac{\OM{k}}{\sqrt{2V\VAREPS{k}}m}
\B{\lambda_{\BD{k}3}\exp(-\ii\VAREPS{k}t+\ii\BD{k}\cdot\BD{r})+\cc},\\
& \lara{\psi(t)|\BD{V}(0,\BD{r})|\psi(t)}=
\lara{\tilde{\psi}(t)|\BD{V}(0,\BD{r})|\tilde{\psi}(t)}
-\sum_\KS  \frac{\BD{\eta}(\BD{k},\sigma)}{\sqrt{2V\VAREPS{k}}}
\B{\lambda_\KS\exp(-\ii\VAREPS{k}t+\ii\BD{k}\cdot\BD{r})+\cc},
\label{vvvv}
\end{align}
where the first terms in these expressions represent the static solution from Sec.~\ref{Static_sec}.
For causal reasons, the Proca field must vanish for fixed $t$ and 
sufficiently large $|\BD{r}|$, which  is not evident from the above expressions
(see Secs.~\ref{Point_sec} and \ref{Extended_sec} for an illustration).
Therefore, the static and non-equilibrium 
solutions  differ at all times. Still, for fixed $\BD{r}$ and $t\to\infty$, 
one expects the Proca  field to approach the static value. This expectation can be
investigated  using the Riemann-Lebesgue lemma but  we shall not pursue
this analysis here. Instead, we discuss two examples in which this behavior occurs
and is accompanied by interesting field dynamics unaccounted for 
by  the Riemann-Lebesgue lemma
(Secs.~\ref{Point_sec} and \ref{Extended_sec}).\footnote{In the example
from  Sec.~\ref{Point_sec}, the convergence 
 to the static solution
is not guaranteed by the Riemann-Lebesgue lemma, which illustrates
the limited usefulness of this lemma in the discussed context.}

\subsection{Energy}
\label{Energy_sub}
Standard manipulations reveal that for $t>0$
\be
\lara{\psi(t)|H_1(0;t)-\zeta|\psi(t)}
=-\lara{\psi(t)|H_0|\psi(t)}.
\label{H0H1}
\ee
This implies that 
\be
{\cal E} = \lara{\psi(t)|H(0;t)|\psi(t)}=\zeta,
\label{Hall}
\ee
which is time independent due to $H(0;t>0)=H$; 
$H$ is given by (\ref{Hdiag}).
Comparing (\ref{H0H1B}) with  (\ref{H0H1}),
we observe that the partitioning of the energy into free-field and
interaction contributions is fundamentally different for static
and expanding field configurations.

Time independence of  (\ref{Hall}), the energy of the expanding field
configuration,  can be also seen as a result 
of completely out of phase oscillations of the free-field and interaction 
energies. That these quantities are time dependent, one  sees from 
\be
 \lara{\psi(t)|H_0|\psi(t)}= 2\sum_\KS \VAREPS{k}|\lambda_\KS|^2 - 2
 \sum_\KS\VAREPS{k}\cos(\VAREPS{k}t)|\lambda_\KS|^2
 \label{dcwihu}
\ee
combined with (\ref{H0H1}).  Still, they reach  well-defined asymptotic values 
\begin{align}
\label{Ex00}
&\lim_{t\to\infty} \lara{\psi(t)|H_0|\psi(t)}  =  2\sum_\KS
\VAREPS{k}|\lambda_\KS|^2,\\
&\lim_{t\to\infty}   \lara{\psi(t)|H_1(0;t)-\zeta|\psi(t)}=
-2\sum_\KS \VAREPS{k}|\lambda_\KS|^2
\label{Ex11}
\end{align}
when  the right-hand sides of these equations are finite
(see  (\ref{RLL}) and the discussion in Appendix \ref{RL_app}).
Comparing these results to their static counterparts,
we see that (\ref{h0h1a}) and (\ref{Ex00})
differ fundamentally  while (\ref{h0h1b})
and (\ref{Ex11}) agree.

Finally, at the risk of stating the obvious,
we observe that the difference between the energies of the expanding and
static fields is positive, which is seen from (\ref{tildeE}) 
and (\ref{Hall}).

\subsection{Energy fluctuations}
\label{Fluctuations_sub}

Energy fluctuations in the state $|\psi(t)\ra$ can be characterized 
for $t>0$ by central moments
\be
\label{dlk}
\lara{(H-\zeta)^n}\equiv\lara{\psi(t)|(H(0;t)-\zeta)^n|\psi(t)},
\ee
which are time independent due to the time independence 
of the Hamiltonian:  $H(0;t>0)=H$ with  $H$  given by
(\ref{Hdiag}). It turns out that 
\begin{align}
\label{meflrmfkler}
&\lara{H-\zeta}=0, \ \lara{(H-\zeta)^2}=h_2, \ \lara{(H-\zeta)^3}=h_3, \ 
\lara{(H-\zeta)^4}=h_4+3 h_2^2,\\
&h_n = \sum_\KS \VAREPS{k}^n |\lambda_\KS|^2.
\label{hnn}
\end{align}

These results can be obtained in  different  ways.
For example,  one may employ 
\be
\lara{(H-\zeta)^n}=\left.\ii^n\frac{d^n}{ds^n}{\cal F}(s)\right|_{s=0},
\ee
where ${\cal F}(s)=\lara{\psi(t)|\exp\B{-\ii (H-\zeta)s}|\psi(t)}
=\exp(\ii\zeta s)\lara{\psi(t)|\psi(t+s)}$, and proceed as in \cite{BDmag}
taking advantage of the fact that the survival amplitude 
$\lara{\psi(t)|\psi(t+s)}$ can be 
exactly obtained in the discussed problem.

Finally, we note that the standard deviation of the 
energy  reads
\be
\sigma_{\cal E}=\sqrt{h_2}.
\label{sigE}
\ee

\subsection{Overlap}
\label{Overlap_sub}
The overlap between states representing static and expanding fields reads
\be
\lara{\tilde{\psi}(t)|\psi(t)} = \exp(\ii\tilde{\cal E}t-\ii\zeta t)
\prod_\KS \exp\B{\ii\text{Im}\Phi_\KS(t)}
 \lara{\lambda_\KS|\alpha_\KS(t)\exp(-\ii\VAREPS{k}t)}
\label{FFf}
\ee
for $t>0$. This quantity is time independent.
While this feature is not apparent  from 
(\ref{FFf}),  standard 
manipulations akin to those performed  in \cite{BDmag} confirm 
both this prediction and the fact that 
$\lara{\tilde{\psi}(t)|\psi(t)}=\lara{\tilde{0}|0}$.
In the end, we arrive at 
\be
\lara{\tilde{\psi}(t)|\psi(t)}=\exp\B{ -\frac{\tilde{\cal N}}{2} },
\label{Ovee}
\ee
which implies  that the overlap is  strictly smaller than unity.
Consequently, the static and expanding states are (perpetually) distinct.

\subsection{Vector-boson number}
\label{Photon_sec}

The number of vector bosons  in the expanding field  and its fluctuations 
are characterized for $t\ge0$ by 
\begin{align}
&{\cal N}(t)=\lara{\psi(t)|N|\psi(t)}
=2\sum_\KS |\lambda_\KS|^2   -2\sum_\KS \cos(\VAREPS{k}t)|\lambda_\KS|^2,\\
&\sigma_{\cal N}(t)=\sqrt{{\cal N}(t)},
\end{align}
respectively. Moreover, using (\ref{phn}),  (\ref{RLL}), and Appendix
\ref{RL_app}, we find that 
\be
\lim_{t\to\infty}{\cal N}(t)=2\tilde{\cal N}
\label{Ntin}
\ee
for  finite $\tilde{\cal N}$.
The difference between (\ref{Ntin}) and (\ref{phn}), 
$\tilde{\cal N}$, can be viewed  as the number of excitations 
generated by the sudden creation of an external current.

\section{Point charge}
\label{Point_sec}
We  employ in this section the point-charge model.
This  simplification allows us to obtain an exact closed-form solution
describing  field dynamics following charge creation.
Such  a solution facilitates the analytical discussion of  the
causal propagation of the  Proca field 
and its  periodically-modulated algebraic approach 
toward the equilibrium value.
The drawback of the point-charge model is that 
some observables introduced in previous sections--those 
related to the energy and the vector-boson  number--are
divergent in this model. 
This deficiency will be addressed in Sec. \ref{Extended_sec},
where these quantities will be discussed within the extended-charge model.

\subsection{Static Proca field}

We consider  the following external current 
\be
\jjmat(\BD{r})=  (q\delta(\BD{r}),\0),
\label{JMATMAG}
\ee
where $\delta(\BD{r})$ is a three-dimensional delta function.
For this external current 
\be
\lambda_\KS= -\frac{q\OM{k}}{\sqrt{2V}m\VAREPS{k}^{3/2}}\delta_{\sigma3},
\label{lks}
\ee
which via (\ref{Vstat0}) and (\ref{VstatVec}) leads to
\begin{align}
\label{V0stat}
&\lara{\tilde{\psi}(t)|V^0(0,\BD{r})|\tilde{\psi}(t)}=
\frac{q}{4\pi r}\exp(-m r),\\
&\lara{\tilde{\psi}(t)|\BD{V}(0,\BD{r})|\tilde{\psi}(t)}
=\0,
\label{Vstat}
\end{align}
where $r=|\BD{r}|$. 
These  results reproduce the well-known 
classical solution of Proca theory
describing the charge $q$ whose field is 
 screened over distances much greater than 
$1/m$ \cite{Nieto_RMP1971,Gillies2005,Nieto_RMP2010}.

\begin{figure}[t]
\includegraphics[width=\pref\columnwidth,clip=true]{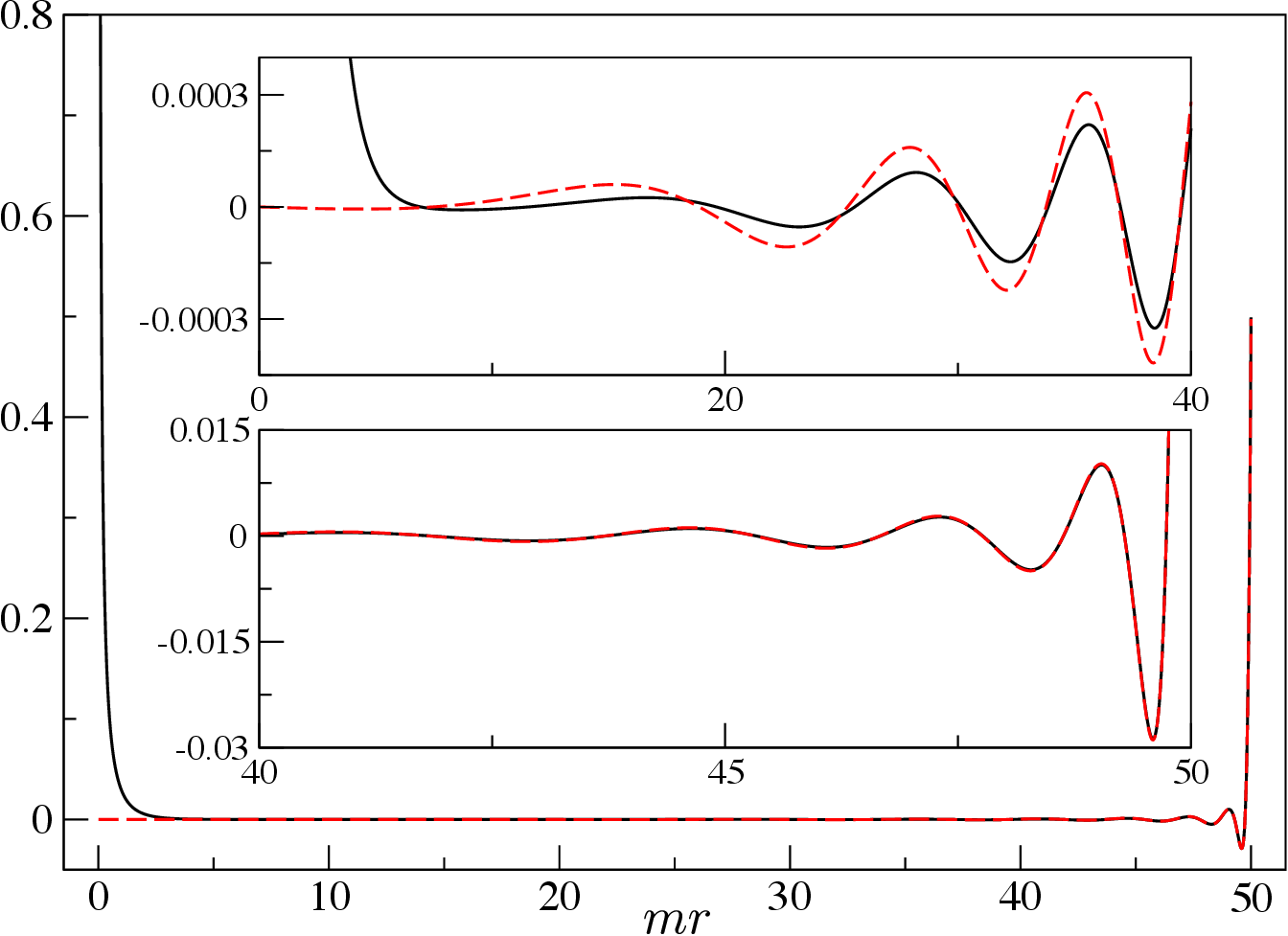}
\caption{The black solid and red dashed lines show
$\lara{\psi(t)|V^0(0,\BD{r};t)|\psi(t)}/(qm)$  and 
$\rhat\cdot\lara{\psi(t)|\BD{V}(0,\BD{r})|\psi(t)}/(qm)$, respectively, for 
$mt=50$. 
They are obtained from (\ref{Ins}) and (\ref{Insvec}).
The insets display selected regions  of the main plot. In the lower inset,
the  black solid and red dashed  lines  are
practically indistinguishable on the plotted scale.
Note that the ranges of the vertical axes in the insets differ by 
two orders of magnitude. 
Numerical calculations  for this and all other plots have been 
performed using \cite{Mathematica143}.
}
\label{point_fig}
\end{figure}

\subsection{Expanding  Proca field}
\label{Expanding_sec}

We  assume now that 
\be
\jjmat(t,\BD{r})=  \B{q\delta(\BD{r})\theta(t),\0},
\label{JmatMAGt}
\ee
and provide below relevant expressions for $t>0$
\begin{align}
\label{V0tr}
& \lara{\psi(t)|V^0(0,\BD{r};t)|\psi(t)}=-q\int_0^t dt'\Delta(t',\BD{r})
-\frac{q}{m^2}\partial_t\Delta(t,\BD{r}),\\
&\lara{\psi(t)|\BD{V}(0,\BD{r})|\psi(t)} = 
\frac{q}{m^2}\bnabla\Delta(t,\BD{r}),
\label{Vvtr}
\end{align}
where the  Pauli-Jordan function is defined via
\be
\Delta(t,\BD{r})\equiv -\ii\int \frac{d^3k}{(2\pi)^3}
\frac{\exp(-\ii\VAREPS{k}t+\ii\BD{k}\cdot\BD{r})}{2\VAREPS{k}}+\cc
\ee
Using \cite{Scharf}, we write it for $t>0$  as
\be
\label{PJ}
\Delta(t,\BD{r})=  \frac{m}{4\pi}
\frac{\BesselJ_1\B{m\sqrt{t^2-r^2}}}{\sqrt{t^2-r^2}}\theta\B{t-r}   -  \frac{\delta(t-r)}{4\pi t},
\ee
where $\BesselJ_n$ is  the Bessel function of the first kind of order $n$ and
$\delta(t-r)$ is   a one-dimensional delta function.
The derivation of (\ref{V0tr}) and (\ref{Vvtr}) is discussed in Appendix \ref{Integrals_app};
see Appendix \ref{Solution_app} for the  self-consistency check of these
results.
Equations  (\ref{V0tr}) and (\ref{Vvtr}) can be separately discussed in three regions.

{\bf At the shockwave front}.
As is evident from the above results, 
expressions for  
$\lara{\psi(t)|V^0(0,\BD{r};t)|\psi(t)}$
and $\lara{\psi(t)|\BD{V}(0,\BD{r})|\psi(t)}$
involve distributions, $\theta(t-r)$ and  $\delta(t-r)$,
describing spatially-singular structures propagating at the speed of light. 
Moreover, as will be clear from the following discussion,
the discussed Proca field  undergoes self-steepening 
dynamics for $r$ approaching $t$ from below. In view of these features,  we say that 
there is a shockwave whose  front is    located at $r=t$. 
We attribute  the  appearance of this  shockwave structure
 to the singular initial conditions  (see the discussion
 in Sec. \ref{Extended_sec} for supporting evidence). 
Finally, we note that the propagation velocity equal to the speed of light is
quite interesting given that we are dealing with a relativistic theory of massive particles
(see \cite{BDPeriodic1} for a somewhat related  example of luminal 
propagation in the Proca theory).

{\bf Outside the area enclosed by the shockwave front}. For $r>t$,
we have 
$\lara{\psi(t)|V^0(0,\BD{r};t)|\psi(t)}=0$ 
and $\lara{\psi(t)|\BD{V}(0,\BD{r})|\psi(t)}=\0$. 
This is immediately evident from the fact that 
$\Delta(t,\BD{r})$ vanishes for spacelike $(t,\BD{r})$. 
This feature is expected because the field undergoes 
causal evolution in a relativistic theory.

{\bf Inside the area enclosed by the shockwave front}. We find for $r<t$
\begin{subequations}
\begin{align}
\label{InsA}
\lara{\psi(t)|V^0(0,\BD{r};t)|\psi(t)} &=  \frac{q}{4\pi r}\exp(-m r) 
+\frac{qm}{4\pi}
\int_{m t}^\infty dx \frac{\BesselJ_1\B{\sqrt{x^2-(m r)^2}}}{\sqrt{x^2-(m r)^2}}\\
& 
+\frac{q t}{4\pi m}\BB{
\frac{2\BesselJ_1\B{m\sqrt{t^2-r^2}}}{(t^2-r^2)^{3/2}}
- \frac{m\BesselJ_0\B{m\sqrt{t^2-r^2}}}{t^2-r^2}
},
\label{InsB}
\end{align}
\label{Ins}%
\end{subequations}
\be
\lara{\psi(t)|\BD{V}(0,\BD{r})|\psi(t)}=\frac{q\BD{r}}{4\pi m}
\B{\frac{2\BesselJ_1\B{m\sqrt{t^2-r^2}}}{(t^2-r^2)^{3/2}}
-\frac{m\BesselJ_0\B{m\sqrt{t^2-r^2}}}{t^2-r^2}
},
\label{Insvec}
\ee
where (\ref{InsA}) and (\ref{InsB}) originate  from the first and second 
term in (\ref{V0tr}), respectively. The derivation of (\ref{InsA}) is discussed 
in Appendix \ref{Derivation_app} while the  derivation of (\ref{InsB}) and
(\ref{Insvec}) 
follows from standard Bessel function identities.
These results are illustrated in Fig. \ref{point_fig}, where 
it is evident that (\ref{Ins}) and the scalar product 
of $\rhat$ and (\ref{Insvec})
closely overlap
for $r\to t^-$. 
This feature appears to be associated with
the following observations. 
In the discussed limit, (i) (\ref{InsA})
is equal to $q/(4\pi t)$, which follows from (\ref{edlk}); 
 (ii) (\ref{InsB}) is equal to $qm^2t/(32\pi)$.
Consequently, the ratio of the results reported in cases (i) and (ii)
 is $8/(mt)^2$, which is much smaller  than unity for $mt=50$ 
used in Fig.~\ref{point_fig}. Thereby, for   $r\to t^-$ and $mt\gg1$, 
(\ref{Ins}) can be approximated by (\ref{InsB}), which 
matches the scalar product of $\rhat$ and (\ref{Insvec})
for $r\to t^-$.

\begin{figure}[t]
\includegraphics[width=\pref\columnwidth,clip=true]{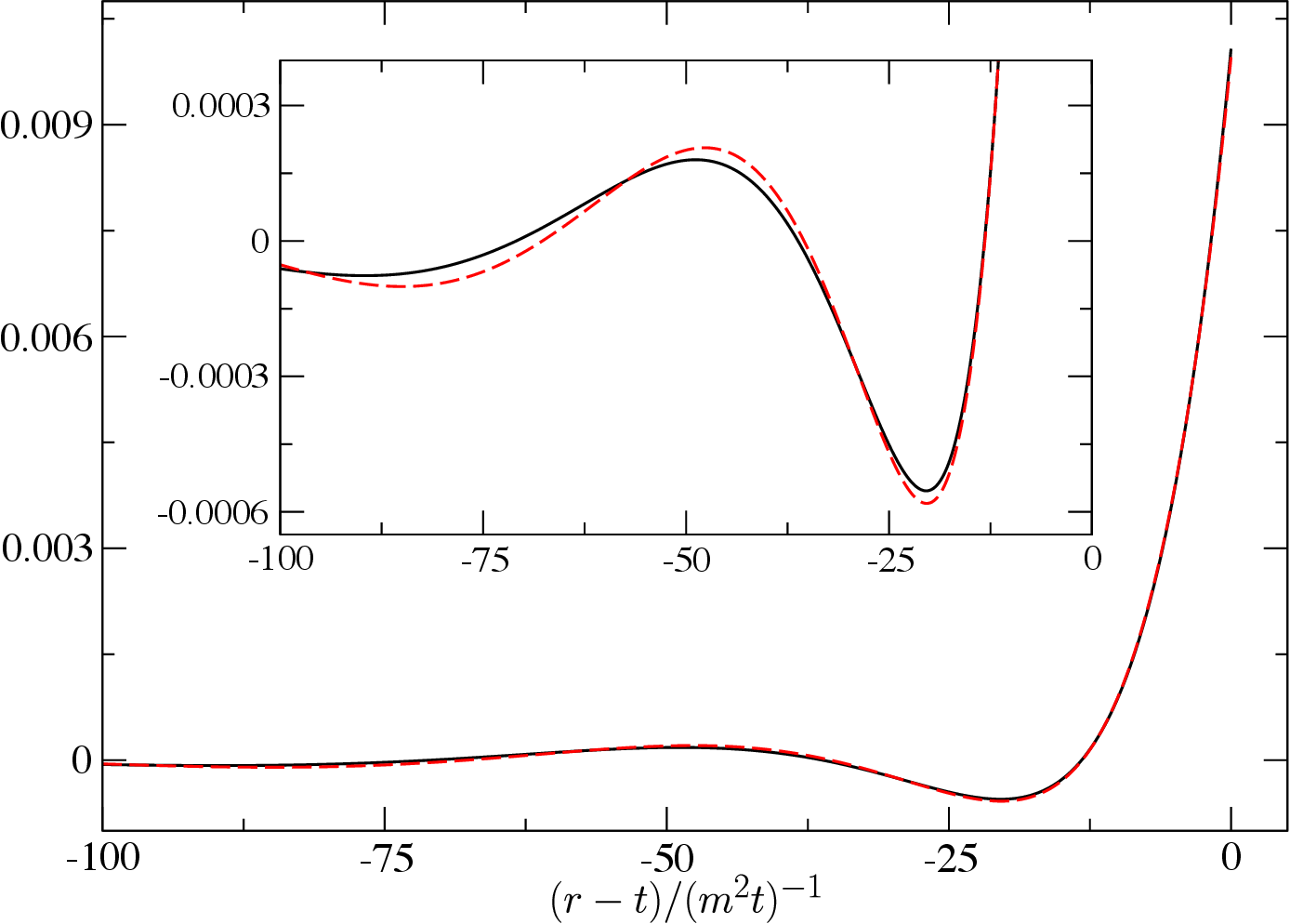}
\caption{
The black solid and red dashed lines  show
$\lara{\psi(t)| V^0(0,\BD{r};t)|\psi(t)}/(qm^2t)$ in the point-charge
model for $mt=25$ and $mt=100$,
respectively.  The  lines closely overlap on the right-hand side of the plot.
The inset displays a selected region of the main plot, highlighting the differences
between the data for
$mt=25$ and $mt=100$ (on the plotted scale, there is practically no difference
between the data for $mt=100$ and $mt=50000$).
A very similar figure  is obtained for
$\protect{\rhat\cdot\lara{\psi(t)| \BD{V}(0,\BD{r})|\psi(t)}/(qm^2  t)}$ and
as such it is not shown.
}
\label{ansatz_fig}
\end{figure}

Then, we observe that the derivative of (\ref{InsB}) with respect to $r$
is given by $q m^4 t^2/(192\pi)$ at $r\to t^-$. This result,
together with the one reported in point (ii), suggests the following ansatz
for $\lara{\psi(t)|V^0(0,\BD{r};t)|\psi(t)}$ and
$\rhat\cdot\lara{\psi(t)|\BD{V}(0,\BD{r})|\psi(t)}$
when $r$ approaches $t$ from below
\be
q m^2 t F\B{\frac{r-t}{(m^2t)^{-1}}},
\label{drtdr}
\ee
where $F$ is a scaling function and $mt\gg1$ is 
assumed.\footnote{The same ansatz may be also  obtained by Taylor expanding (\ref{InsB})
about $r=t$ and retaining, at each order in the expansion, only the dominant
contribution for  $mt\to\infty$.}
This ansatz predicts that as time increases, the quantities under
consideration linearly grow with $t$ at $r=t^-$, whereas the width of
the region over which this value is attained shrinks as $1/(m^2 t)$. 
The validity of (\ref{drtdr})  is illustrated in Fig. \ref{ansatz_fig}.
Regarding the structure of (\ref{drtdr}), we observe that
it is quite unusual in the context of both classical
shockwaves \cite{LandauHydro} and  their quantum
counterparts in many-body cold-atom systems
(see e.g. \cite{BDJPB2004} for an example involving a Fermi gas
and \cite{QuantumShock2020,Olshanii2021} for 
recent examples involving Bose gases).

A different situation is found for $t\to\infty$ at fixed $r$, 
where the lowest-order approximations of 
(\ref{Ins}) and (\ref{Insvec}) yield
\begin{align}
\label{Posc}
&\lara{\psi(t)|V^0(0,\BD{r};t)|\psi(t)}-\frac{q}{4\pi r}\exp(-m r)
\simeq \frac{3q m}{(2\pi)^{3/2}}\frac{\sin\B{m t-\pi/4}}{(mt)^{5/2}},\\
&\rhat\cdot\lara{\psi(t)|\BD{V}(0,\BD{r})|\psi(t)}
\simeq -\frac{qm^2r}{(2\pi)^{3/2}}\frac{\cos(mt-\pi/4)}{(mt)^{5/2}},
\label{PPosc}
\end{align}
whose validity is  illustrated in Fig. \ref{point_osc}. 
The derivation of (\ref{Posc}) is presented in Appendix~\ref{Der_app};
see \cite{Ryzhik} for an asymptotic expansion of the Bessel functions employed in the derivation
of both formulas.
The damped oscillatory behavior encoded in (\ref{Posc}) and (\ref{PPosc}) 
is interesting for three   reasons.

First, although the fact that
(\ref{Posc}) and (\ref{PPosc}) vanish in the long-time 
limit is intuitively plausible--the Proca field locally approaches its 
equilibrium value in the presence of the external source--the appearance of this 
property is not guaranteed by the Riemann–Lebesgue lemma-based argument 
employed in Sec.~\ref{Vector_sub}. This is because
$\lambda_{\BD{k}3}(\OM{k}\to\infty)\propto \OM{k}^{-1/2}$
 does not decay sufficiently fast for the applicability of (\ref{RLL}).

Second, the oscillatory behavior of (\ref{Posc}) and (\ref{PPosc}) contrasts
with the results  reported in \cite{BDmag}, 
where the dynamics  of a massless vector field, after the sudden 
creation of an extended  magnetic dipole,  
was analytically solved. 
In that case, no oscillations are present: 
after the light-speed passage of the shockwave-like  front 
through a given spatial point, the vector field reaches its equilibrium value.
We attribute the appearance of damped oscillations in the Proca theory to 
the presence of massive particles, which (unlike photons) 
can propagate with various  velocities. Since the wavepacket
generated during the charge-creation process contains massive vector bosons 
propagating with 
a broad range of velocities, these particles reach a given spatial point at different
times rather than ``all at once''. Quite interestingly, the period of these damped
oscillations is given by a characteristic time scale associated with the 
Proca theory,
\be
T_\text{osc}= \frac{2\pi}{m}.
\label{Tosc}
\ee

Third, quantum transients are temporary features of matter waves that
arise   before the system reaches 
a stationary regime \cite{TransientsReview}. Thereby,  the discussed damped 
oscillations can be regarded as a quantum transient.

Finally, we emphasize that the above discussion supports 
the view that the non-equilibrium solution studied in this
work represents an expanding field. For the same reasons, 
the same terminology applies to the solution discussed in the next section.

\begin{figure}[t]
\includegraphics[width=\pref\columnwidth,clip=true]{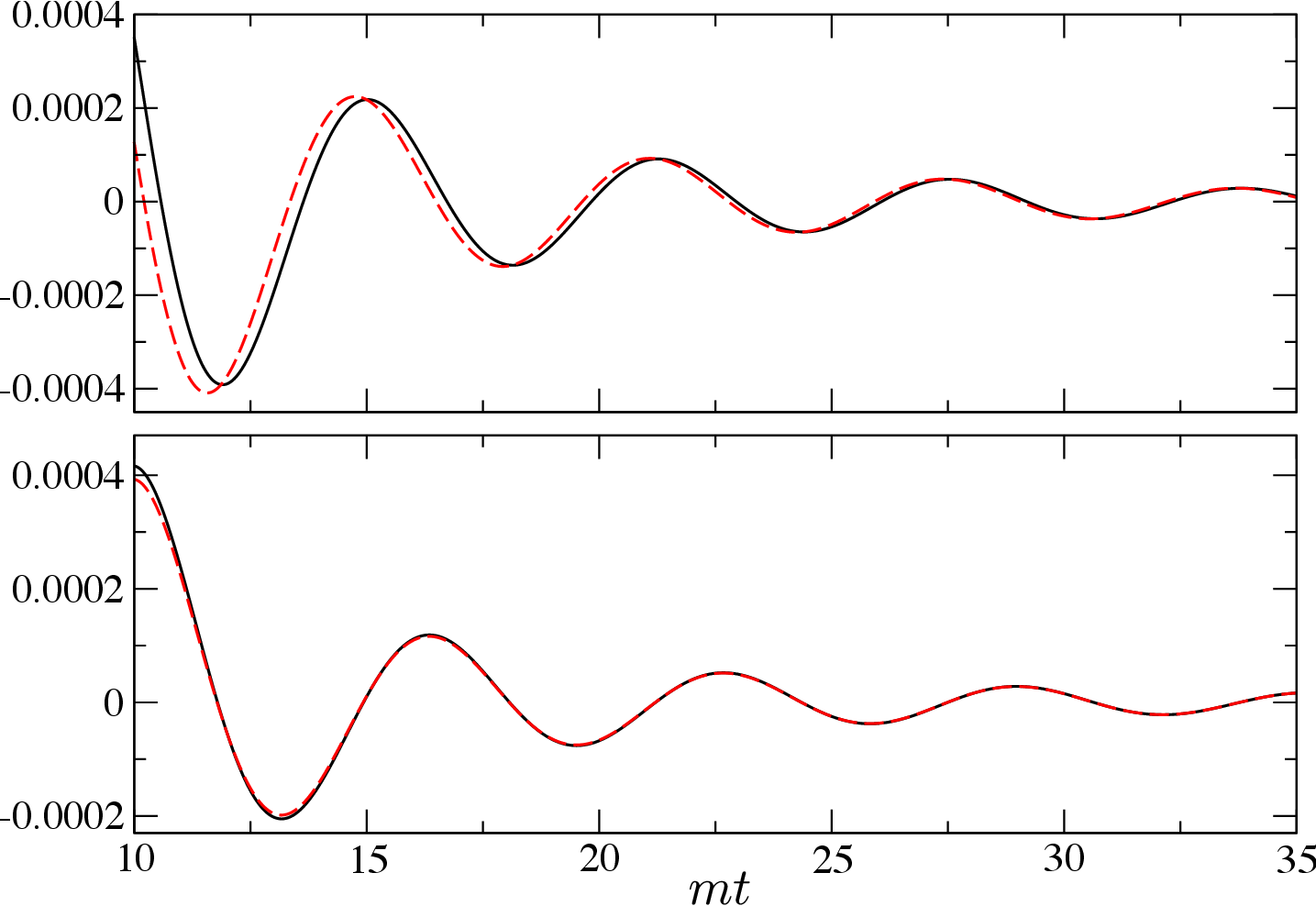}
\caption{
Upper panel: the black solid line shows the left-hand side of (\ref{Posc}) multiplied by 
$1/(qm)$  while the red dashed line shows 
the right-hand side of (\ref{Posc}) multiplied by the same factor.
Lower panel: it  is structured in the same way as the upper panel but
it deals with the left- and right-hand sides of (\ref{PPosc}).
In both panels, $mr=2$ and the black solid and red dashed lines  are
practically indistinguishable on the plotted scale for long-enough times.
}
\label{point_osc}
\end{figure}

\section{Extended charge}
\label{Extended_sec}

We discuss here changes to the point-charge results  when one works
with  the Gaussian formulation of the  extended-charge model, 
\be
\label{jrs}
\jjmat(\BD{r})=\B{q\delta_\epsilon(\BD{r}),\0}, 
\ee
where  
\be
\delta_\epsilon(\BD{r})=
\frac{\exp\B{-r^2/\epsilon^2}}{(\sqrt{\pi}\epsilon)^3}
\label{lkewfkw}
\ee
is  a nascent delta function   so that the point-charge model is recovered for $\epsilon\to0^+$.
For this choice  of the external current,
\be
\lambda_\KS= -\frac{q\OM{k}\exp(-\epsilon^2\OM{k}^2/4)}{\sqrt{2V}m\VAREPS{k}^{3/2}}\delta_{\sigma3}.
\label{lksEX}
\ee
Both  static- and expanding-field calculations rely on this  quantity.
Before discussing our findings, we have the following  general remarks. 

First, we provide exact results for time-independent quantities:
$\cal E$ and  $\sigma_{\cal E}$ derived in Appendix~\ref{IntegralSTAT_E_app}
as well as  $\lara{\tilde{\psi}(t)| V^0(0,\BD{r})  |\tilde{\psi}(t)}$, 
$\tilde{\cal E}$, and $\tilde{\cal N}$  derived in Appendix~\ref{IntegralSTAT_S_app}.
We emphasize  that they   are all valid for arbitrary $\epsilon>0$.

Second, straightforward calculations reveal that the time-dependent quantities of interest are 
given for $t>0$ by 
\begin{align}
\label{v0t}
&\lara{\psi(t)|V^0(0,\BD{r};t)|\psi(t)}=
\lara{\tilde{\psi}(t)|V^0(0,\BD{r})|\tilde{\psi}(t)}
+\frac{q}{2\pi^2m^ 2r}\int_0^\infty d\OM{k}\frac{\OM{k}^3\exp(-\epsilon^2\OM{k}^2/4)}{\VAREPS{k}^2}
\sin(\OM{k}r)
\cos(\VAREPS{k}t),\\
\label{vvect}
&\lara{\psi(t)|\BD{V}(0,\BD{r})|\psi(t)}=\frac{q\rhat}{2\pi^2m^2}
\int_0^\infty
d\OM{k}\frac{\OM{k}^3\exp(-\epsilon^2\OM{k}^2/4)  }{\VAREPS{k}}\SphBesselj_1(\OM{k}r)\sin(\VAREPS{k}t),\\
&\lara{\psi(t)|H_0|\psi(t)}=2({\cal E}-\tilde{\cal E})-\frac{q^2}{2\pi^2m^2}\int_0^\infty d\OM{k}
\frac{\OM{k}^4\exp(-\epsilon^2\OM{k}^2/2) }{\VAREPS{k}^2}\cos(\VAREPS{k}t),\\
&{\cal N}(t)=2\tilde{\cal N}-
\frac{q^2}{2\pi^2m^2}\int_0^\infty d\OM{k}
\frac{\OM{k}^4\exp(-\epsilon^2\OM{k}^2/2) }{\VAREPS{k}^3}\cos(\VAREPS{k}t),
\end{align}
where $\SphBesselj_n$ is the spherical Bessel function of the first kind of order
$n$.\footnote{As a self-consistency check, we have verified that
(\ref{v0t}) and (\ref{vvect}) satisfy the Proca field equations with external 
current (\ref{jrs}).}
The integrals in the above expressions are numerically evaluated, 
whereas the remaining contributions  are obtained via 
(\ref{efw}), (\ref{wwds}),  (\ref{Estt}),  and (\ref{ewuy}).
The numerics
is  done for $m\epsilon=0.01\ll1$ because the  goal of this section  
is to  supplement the discussion presented in Sec.~\ref{Point_sec},
which is necessitated by  the singular nature of the point-charge model.

Third,  we observe that all general results from Sec.~\ref{Problem_sec},
which rely on the applicability of (\ref{RLL}), hold for $\lambda_\KS$ given by (\ref{lksEX}).

\begin{figure}[t]
\includegraphics[width=\pref\columnwidth,clip=true]{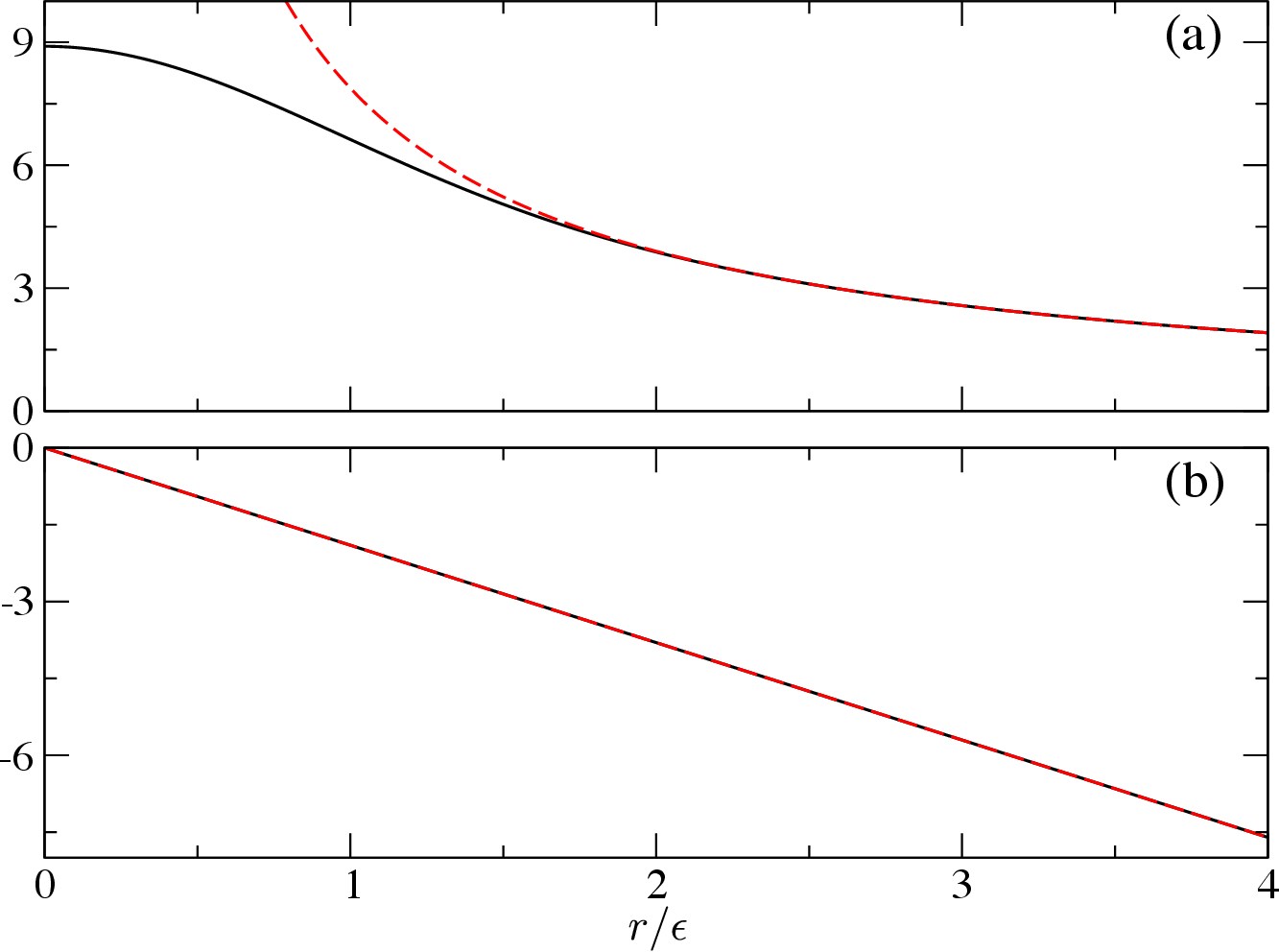}
\caption{Panel (a): the black solid  line shows  
$\lara{\psi(t)| V^0(0,\BD{r};t)|\psi(t)}/(qm)$ in the extended-charge
model  for $m\epsilon=0.01$ and $mt=50$.   
The red dashed line shows the same quantity 
in the point-charge model.
Panel (b): it  is organized in the same way as  panel (a) but
it deals with  
$\protect{\rhat\cdot\lara{\psi(t)| \BD{V}(0,\BD{r})|\psi(t)}\times10^8/(qm)}$; the black solid and 
red dashed lines are
indistinguishable within the resolution of the plotted scale.
}
\label{near_fig}
\end{figure}

\subsection{Static Proca field}

The static Proca field   reads 
\begin{align}
\label{efw}
&\lara{\tilde{\psi}(t)| V^0(0,\BD{r})  |\tilde{\psi}(t)}=
-\frac{q}{8\pi r} \exp\B{ \frac{(m\epsilon)^2}{4}  }
\BB{
 \exp(m r)\erfc\B{\frac{m\epsilon}{2}+\frac{r}{\epsilon}} 
-\exp(-m r)\erfc\B{\frac{m\epsilon}{2}-\frac{r}{\epsilon}}
},\\
&\lara{\tilde{\psi}(t)|\BD{V}(0,\BD{r})  |\tilde{\psi}(t)}=\0,
\label{efwZERO}
\end{align}
where $\erfc(x)=1-2\int_0^x dy\exp(-y^2)/\sqrt{\pi}$ is 
the complementary error function.\footnote{As a self-consistency check, 
we have verified that $(-\Delta+m^2)\lara{\tilde{\psi}(t)| V^0(0,\BD{r})
|\tilde{\psi}(t)}=q\delta_\epsilon(\BD{r})$, which agrees with 
the ``static'' Proca field equations with external current (\ref{jrs}).} 
The same solution appeared in the studies of the Abelian Higgs model
\cite{Forgacs2021}.\footnote{
There appear to be typos in \cite{Forgacs2021} in the definition of
the error function and in the expression for its asymptotic expansion.}
It can be simplified by means of the following 
standard result: $\erfc(x)\simeq \exp(-x^2)/(\sqrt{\pi}x)$
for $x\gg1$. For   $r/\epsilon\gg1+\ME/2$, this leads to 
\be
\lara{\tilde{\psi}(t)| V^0(0,\BD{r})  |\tilde{\psi}(t)}\simeq
\frac{q\exp\B{(m\epsilon)^2/4}}{4\pi r}\exp(-m r),
\ee
where $q\exp\B{(m\epsilon)^2/4}$ 
plays the role of the effective  charge magnitude, which is evident after comparing 
the above result with (\ref{V0stat}). 
 As this  interesting phenomenon 
has already been  observed in \cite{Forgacs2021}, we shall not 
dwell on it. Besides, the main stress in this section is 
on results for $m\epsilon\ll1$, making $\exp\B{(m\epsilon)^2/4}\simeq1$
within the resolution of the plotted data.
For $r/\epsilon\ll1$, one may approximate  $\lara{\tilde{\psi}(t)| V^0(0,\BD{r})  |\tilde{\psi}(t)}$ 
by the first term of the Maclaurin  expansion in $r$ 
\be
\lara{\tilde{\psi}(t)| V^0(0,\BD{r})  |\tilde{\psi}(t)}\simeq
\frac{q}{2\pi^{3/2}\epsilon} - \frac{qm}{4\pi}\exp\B{ \frac{(m\epsilon)^2}{4}}
\erfc\B{\frac{m\epsilon}{2}}.
\label{divv1}
\ee
This regular expression can be contrasted with singular (\ref{V0stat}).

\begin{figure}[t]
\includegraphics[width=\pref\columnwidth,clip=true]{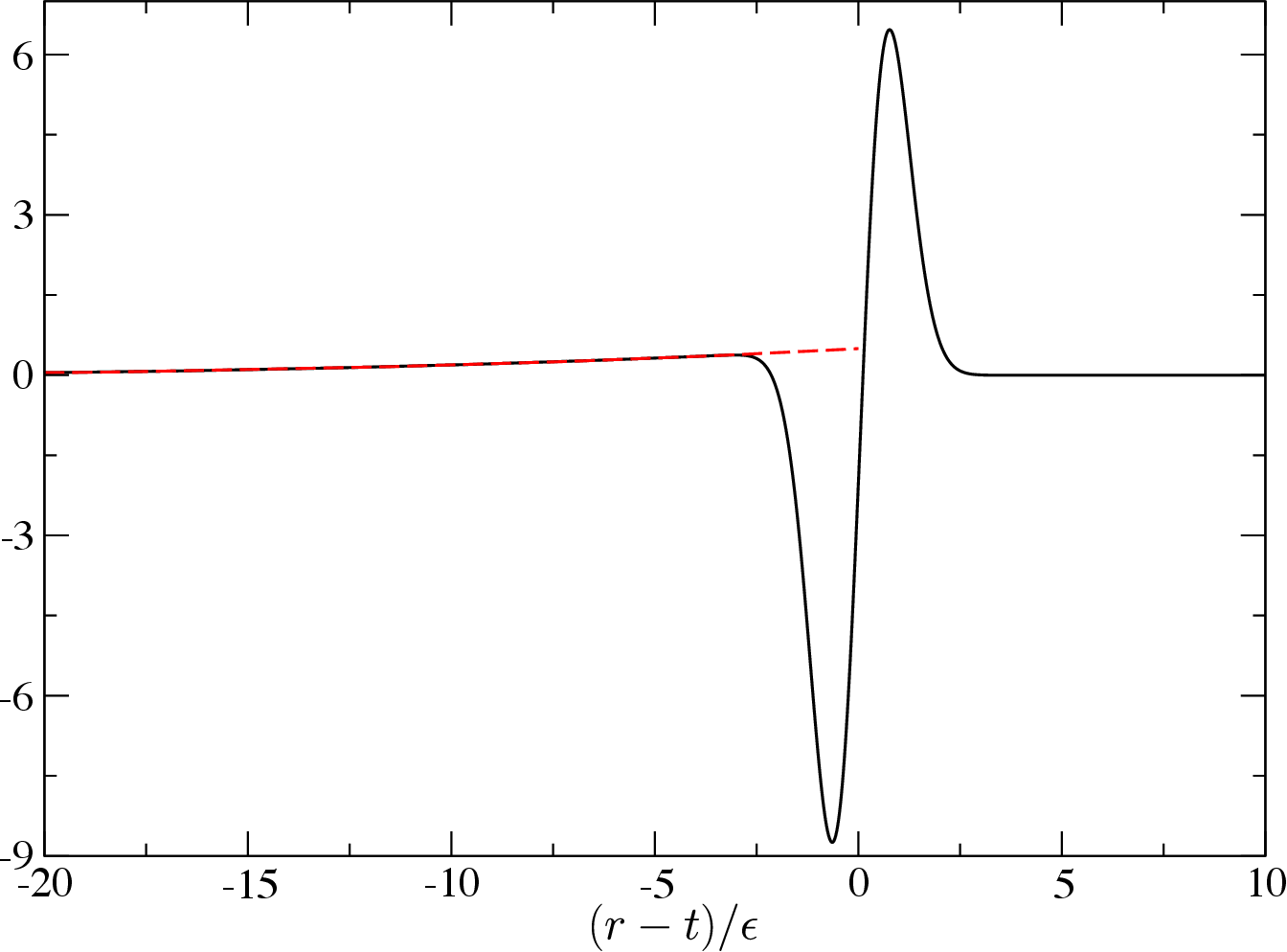}
\caption{The black solid line shows  
$\lara{\psi(t)| V^0(0,\BD{r};t)|\psi(t)}/(qm)$ in the extended-charge
model for $m\epsilon=0.01$ and $mt=50$.  
The red dashed line shows the same quantity 
in the point-charge model for $r<t$.
Note the  close overlap of the black solid and red dashed lines for $(r-t)/\epsilon\lesssim
-1$.
Within the resolution of  the plotted scale,
an indistinguishable plot  is obtained  when 
$\protect{\rhat\cdot\lara{\psi(t)| \BD{V}(0,\BD{r})|\psi(t)}/(qm)}$
is showed 
instead of $\lara{\psi(t)| V^0(0,\BD{r};t)|\psi(t)}/(qm)$.
}
\label{around_fig}
\end{figure}

\subsection{Expanding  Proca field}

In Figs. \ref{near_fig} and  \ref{around_fig}, we
have compared the extended-charge numerics with the  point-charge results
from Fig. \ref{point_fig}. For this purpose, we have chosen  $m\epsilon=0.01$
and $mt=50$. We have found that  
the extended-charge results
closely reproduce the point-charge ones
for $r\gg \epsilon$ and $t-r\gg \epsilon$; 
the two solutions are 
as similar as for large-enough $r/\epsilon$ in 
Fig. \ref{near_fig}a or small-enough $(r-t)/\epsilon$ 
in Fig. \ref{around_fig}. 
The only significant differences between 
them  occur  near $r=0^+$ and around $r=t$.
Three remarks are in order now.

First, discrepancies  near $r=0^+$ are anticipated because 
the extended- and point-charge models markedly differ for $r\lesssim\epsilon$.
They are seen in Fig. \ref{near_fig}a but, somewhat surprisingly, 
 are absent in Fig. \ref{near_fig}b.

Second, whereas the point-charge model exhibits a sharp transition to the field-free region, 
the extended-charge model exhibits a smooth crossover in the vicinity of $r=t$ (Fig. \ref{around_fig}).
This is the
consequence of the extended spatial support of the switched-on external
current,
the propagation of the Proca field  is  causal
in the extended-charge model. 

\begin{figure}[t]
\includegraphics[width=\pref\columnwidth,clip=true]{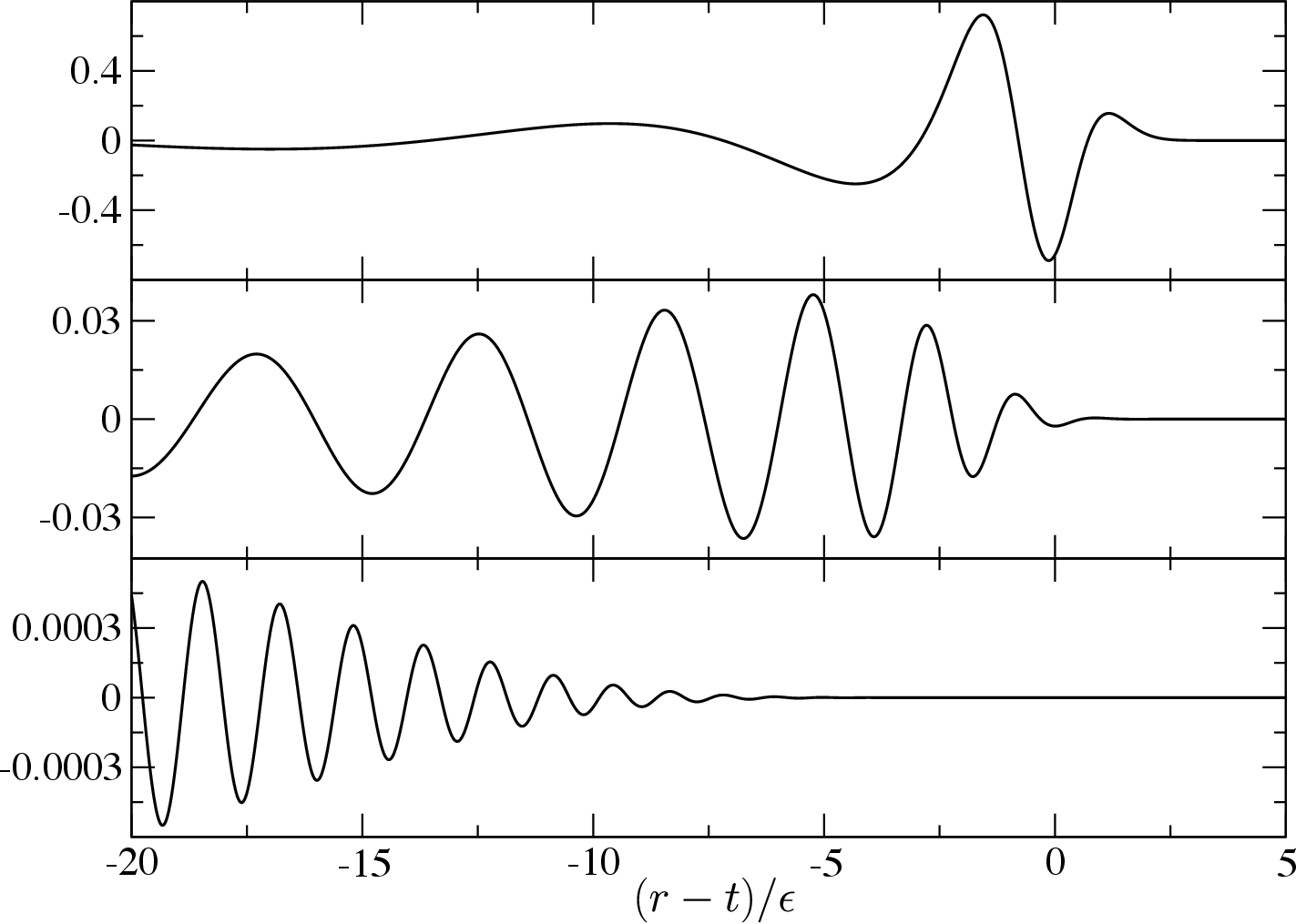}
\caption{From  top to bottom, panels show
$\lara{\psi(t)| V^0(0,\BD{r};t)|\psi(t)}/(qm)$ for 
$m\epsilon=0.01$ and $mt=500$, $5000$, and $50000$, respectively. 
Within the resolution of  the plotted scale,
this  plot  is indistinguishable from that obtained using 
$\protect{\rhat\cdot\lara{\psi(t)| \BD{V}(0,\BD{r})|\psi(t)}/(qm)}$
instead of $\lara{\psi(t)| V^0(0,\BD{r};t)|\psi(t)}/(qm)$.
}
\label{around_ext_fig}
\end{figure}

Third, the long-time dynamics near $r=t$
is presented in Fig.~\ref{around_ext_fig}.
We observe there the temporal decay of the amplitude and
the absence of self-steepening dynamics
akin to that contained in  ansatz (\ref{drtdr}).
Accordingly, the results reported in
Fig.~\ref{around_ext_fig} contrast with those discussed in
Sec.~\ref{Point_sec}. This observation
supports the view that the shockwave features of the solution
presented in Sec.~\ref{Point_sec} stem from singular initial conditions.
The moving structure that rapidly varies  in space,
shown in Figs.~\ref{around_fig} and \ref{around_ext_fig}, will be referred to as
the propagating  front.

\begin{figure}[t]
\includegraphics[width=\pref\columnwidth,clip=true]{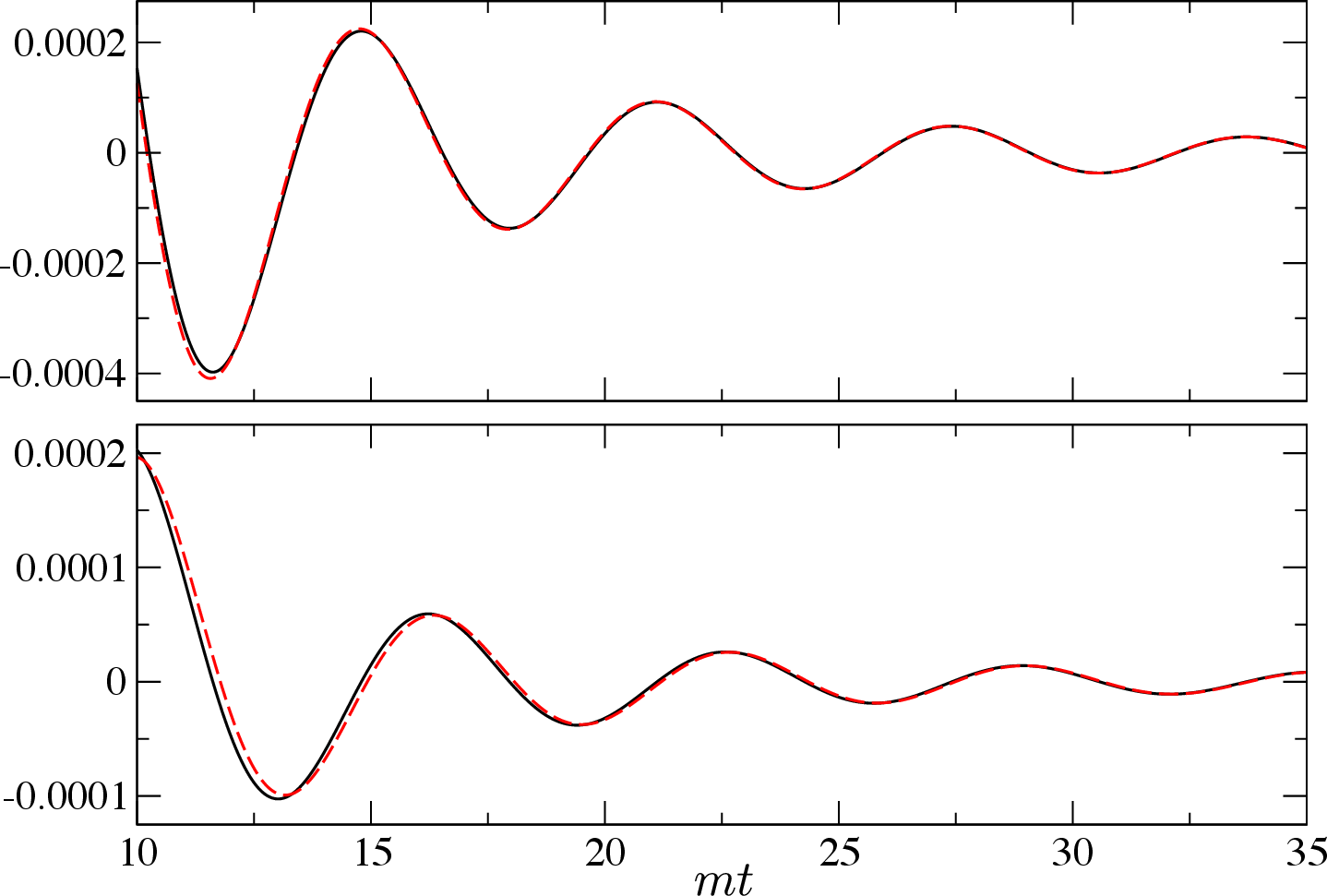}
\caption{Upper panel: the black solid line shows 
$\lara{\psi(t)|V^0(0,\BD{r};t)|\psi(t)}/(qm)-\lara{\tilde{\psi}(t)|V^0(0,\BD{r})|\tilde{\psi}(t)}/(qm)$
in the extended-charge model 
while the red dashed line shows the right-hand side of (\ref{Posc}) multiplied by
$1/(qm)$. 
Lower panel: the black solid line shows 
$\protect{\rhat\cdot\lara{\psi(t)|\BD{V}(0,\BD{r})|\psi(t)}\times10^3/(qm)}$
in the extended-charge model
while the red dashed line shows the right-hand side of (\ref{PPosc}) multiplied by 
$10^3/(qm)$.  In both panels, $\ME=0.01$,  $r=\epsilon/10$, and 
the black solid and red dashed  lines  are
practically indistinguishable on the plotted scale for long-enough times.
}
\label{ext_osc_fig}
\end{figure}

Then,  we  note that in the area swept by the  propagating  
front, the Proca 
field must ``relax'' to the equilibrium value (Sec.~\ref{Vector_sub}). 
Inspired by the  point-charge results, we have 
compared
$\lara{\psi(t)|V^0(0,\BD{r};t)|\psi(t)}-\lara{\tilde{\psi}(t)|V^0(0,\BD{r})|\tilde{\psi}(t)}$
and 
$\protect{\rhat\cdot\lara{\psi(t)|\BD{V}(0,\BD{r})|\psi(t)}}$--both
computed in the extended-charge model--with the right-hand sides of 
(\ref{Posc}) and (\ref{PPosc}), respectively.
This has been done for $\ME=0.01$ and $mr$ equal to either $2$ or $m\epsilon/10$.
In the former case, the results look just as in Fig.~\ref{point_osc} within
the resolution of the plotted scale. In the latter case, the results are shown 
in Fig.~\ref{ext_osc_fig}.

Finally, we note that the expanding solution discussed in this work 
shares certain features with the incoming and outgoing solutions 
considered in our earlier studies (see e.g. \cite{BDPeriodic1}).

\subsection{Energy, vector-boson number, and their fluctuations}

\begin{figure}[t]
\includegraphics[width=\pref\columnwidth,clip=true]{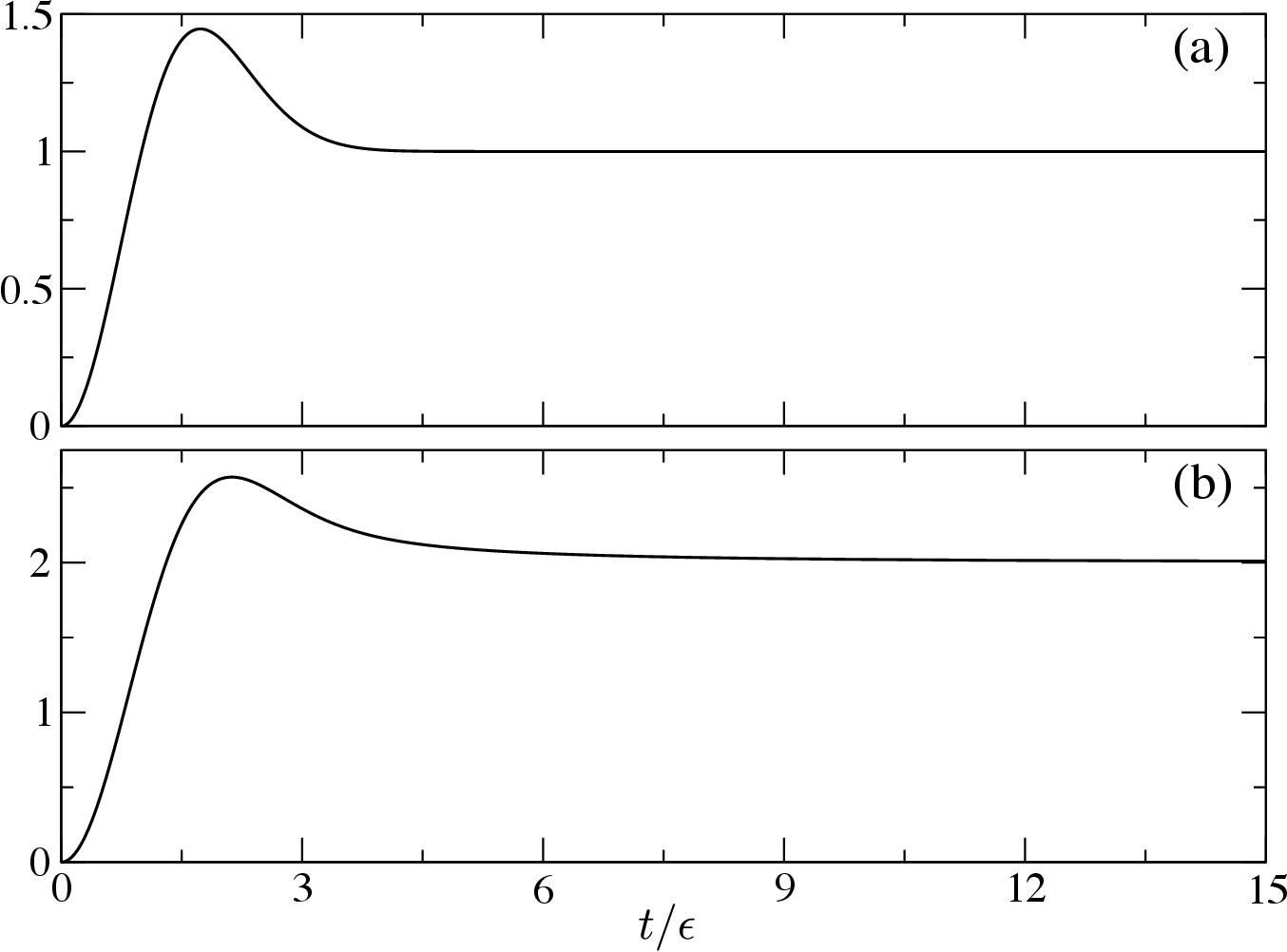}
\caption{Panels  (a) and (b): $\lara{\psi(t)|H_0|\psi(t)}/(2{\cal E}-2\tilde{\cal E})$
and ${\cal N}(t)/\tilde{\cal N}$, respectively.   
The results presented in both panels 
have been obtained within  the extended-charge model for $m\epsilon=0.01$.
}
\label{photon_H0_fig}
\end{figure}

The energy of the static and expanding fields  is 
given by
\begin{align}
\label{wwds}
&\tilde{\cal E} =
\frac{q^2}{2(2\pi)^{3/2}\epsilon} \BB{
1-\sqrt{\frac{\pi}{2}}m\epsilon
\exp\B{\frac{(m\epsilon)^2}{2}}\erfc\B{\frac{m\epsilon}{\sqrt{2}}}},\\
&{\cal E}= \frac{q^2}{2(2\pi)^{3/2}(\ME)^2\epsilon},
\label{Estt}
\end{align}
respectively.
These quantities determine the free-field energy of the expanding solution
for  $t\to\infty$, which is  given by $2{\cal E}-2\tilde{\cal E}$.
This conclusion  follows from
(\ref{Ex00}) expressed in terms of  (\ref{tildeE}) and (\ref{Hall}).
We show the dynamical approach to the discussed limit in Fig. \ref{photon_H0_fig}a.
We observe that one may similarly discuss the dynamics of the 
interaction energy of the expanding field because the sum of 
the free-field and interaction energies of the expanding field   is constant in time (\ref{Hall}).
We also observe that 
$\tilde{\cal E}/{\cal E}\simeq (m\epsilon)^2$ for  $\epsilon\to0^+$,
the energy of the static field is small as compared to the energy of the expanding field.

The standard deviation  of the energy of the static field  
vanishes,  whereas the 
standard deviation of the energy of the expanding field   
is given by
\be
\sigma_{\cal E}
=\frac{q}{4\pi\epsilon}\exp\B{\frac{(m\epsilon)^2}{8}}
\sqrt{(m\epsilon)^2 \BesselK_0\B{ \frac{(m\epsilon)^2}{4}  } + 
\B{2-(m\epsilon)^2}\BesselK_1\B{ \frac{(m\epsilon)^2}{4}  }},
\label{dew}
\ee
where $\BesselK_n$ is the modified Bessel function of the second kind of order
$n$. We observe that $\sigma_{\cal E}/{\cal E} \propto\ME$ for $\epsilon\to0^+$,
the energy fluctuations  of the expanding field are relatively
small  in this limit.

The number of vector bosons  in the static field configuration 
reads 
\begin{align}
\tilde{\cal N}= 
\frac{q^2}{(4\pi)^2}\exp\B{\frac{(m\epsilon)^2}{4}}
\BB{\B{1+(m\epsilon)^2} \BesselK_1\B{ \frac{(m\epsilon)^2}{4}  } -
\B{3+(m\epsilon)^2}\BesselK_0\B{ \frac{(m\epsilon)^2}{4}  }}.
\label{ewuy}
\end{align}
On the one hand, this quantity determines the overlap of expanding- and static-field 
wavefunctions  via (\ref{Ovee}). On the other hand, it  determines via (\ref{Ntin})
the number of vector bosons in the expanding field 
in the long-time limit. Prediction (\ref{Ntin}) is illustrated in Fig.
\ref{photon_H0_fig}b, where the dynamics of the vector-boson number of the 
expanding field is presented. 
Moreover, 
$\tilde{\sigma}_{\cal N}/\tilde{\cal N}=1/\sqrt{\tilde{\cal N}}$
and 
$\lim_{t\to\infty}\sigma_{\cal N}(t)/{\cal N}(t)=1/\sqrt{2\tilde{\cal N}}$
are  $\propto \ME$ for $\epsilon\to0^+$, 
the fluctuations in the vector-boson number 
are relatively small for 
the static and expanding fields.
Finally, we comment on two limiting cases.

First, for $\epsilon\to0^+$ at fixed $m$,
the static and expanding field configurations
are characterized by 
\be
\tilde{\cal E} \simeq \frac{q^2}{2(2\pi)^{3/2}\epsilon},\
\tilde{\sigma}_{\cal E}=0, \
\tilde{\cal N}  \simeq 
\frac{q^2}{(2\pi m\epsilon)^2},\
\tilde{\sigma}_{\cal N}  \simeq \frac{q}{2\pi m\epsilon} 
\label{fewnkl}
\ee
and 
\be
{\cal E} = \frac{q^2}{2(2\pi)^{3/2}(m\epsilon)^2\epsilon}, \ 
\sigma_{\cal E}  \simeq  \frac{q}{\sqrt{2}\pi (m\epsilon)\epsilon},
\ {\cal N}(t\to\infty)  \simeq \frac{q^2}{2(\pi m\epsilon)^2},\
\sigma_{\cal N}(t\to\infty) \simeq \frac{q}{\sqrt{2}\pi m\epsilon},
\label{erj}
\ee
respectively. These  results, except for $\tilde{\sigma}_{\cal E}$,
exhibit ultraviolet singularity 
associated with the point-charge limit. Note that it follows from (\ref{lksEX}) 
that $1/\epsilon$ plays the role of the ultraviolet cutoff.

Second, for $m\to0^+$ at fixed $\epsilon$, (\ref{fewnkl}) and (\ref{erj}) still hold.
The situation now is quite intriguing because this limit takes us
towards the massless theory and Proca field operator (\ref{A0r}) is not well defined 
at $m=0$ because longitudinal polarization vector (\ref{k3}) 
scales as $1/m$. Still, it can be  verified 
that the energy of the static field configuration, described by the classical 
Maxwell's
equations with external current (\ref{jrs}), is the same as $\tilde{\cal E}$ given in (\ref{fewnkl}).
The finiteness of $\tilde{\cal E}$ for  $m\to0^+$ results from the  cancellation between 
the singularity induced by the longitudinal polarization vector, which produces the factor 
$1/m^2$  in
$|\lambda_{\BD{k}3}|^2$, and the singularity arising from the factor $1/m^2$ in  (\ref{zetA}).
No such cancellations occur for other  quantities listed in (\ref{fewnkl}) and (\ref{erj}), 
which are divergent except for $\tilde{\sigma}_{\cal E}$.
Moreover, of particular interest is
\be
{\cal E}-\tilde{\cal E}\simeq\B{\frac{1}{(m\epsilon)^2}-1}\frac{q^2}{2(2\pi)^{3/2}\epsilon},
\ee
which can be interpreted as an energetic cost associated with the sudden
creation of a charge in the Proca theory. 
The divergence of this quantity in the considered limit may be viewed 
as the manifestation of the system's resistance 
to the process forbidden in the Maxwell theory (see
\cite{Nieto_RMP2010} for the  same  reflection reached  from a different
angle).

\section{Summary}
\label{Summary_sec}

The primary focus of this work has been the discussion of a quantum state 
encoding the Proca field of a newly created charge.
This state, when the charge is modeled by an external current,
is represented by a coherent state. It describes an expanding Proca 
field that causally
propagates information about the charge creation through space.
We have studied the physical properties encoded in this state after the
sudden onset of an otherwise static external
current--the energy, the vector-boson number, and their
fluctuations--without assuming a specific spatial
model of the charge distribution.
These considerations have subsequently been illustrated using explicit
examples based on point (Gaussian) charge distribution models,
where the expansion of the Proca field is accompanied by a
shockwave (propagating) front. 
In the region swept by this front,
there is transient quantum behavior eventually leading to
the local establishment of a static Proca field.
Outside this region, the discussed Proca field is absent for causal reasons.
A state representing an expanding Proca field has 
been contrasted with the one describing the corresponding static Proca field.
It has been shown that these states always remain distinct and that
their overlap decays exponentially with the
number of vector bosons building  the static Proca field.

The question now is what physical systems can be considered
in the context of an expanding Proca field,
should a source  producing this field  be created.

\begin{table}[t]
\centering
\begin{tabular}{c|c|c|c|c|c}
 &  $m$ [kg] & $\tau$ [s] & $c\tau$ [m] &  $T_\text{osc}$ [s] & $\tau/T_\text{osc}$   \\
\hline
 \ $\rho$ \ & $1.4\times10^{-27}$ & $4.5\times10^{-24}$ & $1.3\times10^{-15}$ & $5.3\times10^{-24}$ &  0.83 \\
\hline
\ $\omega$ \  & $1.4\times10^{-27}$ & $7.6\times10^{-23}$ & $2.3\times10^{-14}$ & $5.3\times10^{-24}$ &  14  \\
\hline
\ $\gamma$ \  & $10^{-49}\div10^{-54}$ &  &   &  $7.4\times10^{-2}\div7.4\times10^{3}$ &  
\end{tabular}
\caption{
Parameters characterizing  $\rho(770)^0$ and $\omega(782)$ mesons
as well as a hypothetical massive photon of unknown lifetime. 
The symbol $\tau$ denotes a mean lifetime, whereas $T_\text{osc}$
denotes the period of damped oscillations in the  quantum 
transient regime  (\ref{Tosc}). For the photon mass, we substitute the 
standard upper bounds. Accordingly, the corresponding lower bounds 
on $T_\text{osc}$ are reported for the photon. SI units are used in this table.
}
\label{ro_tab}
\end{table}

As far as Standard Model particles are concerned, we note that 
its massive vector bosons are unstable. Still, one may ask how 
an expanding Proca field behaves on the time scale of the order of
the mean particle lifetime. To get a feeling for what we are
dealing with here, we have collected the parameters relevant 
for  $\rho$ and $\omega$ mesons in Table \ref{ro_tab}.
We see there that on this time scale, the expanding field reaches 
distances of the order of a few proton radii and does not evolve
long enough to reach an equilibrium value at any  point
traversed by the shockwave (propagating) front. The former remark follows
from comparing the middle column of Table \ref{ro_tab} with the 
values reported in \cite{ProtonScience2013}. The latter remark follows
from the values reported in the last column of Table \ref{ro_tab},
indicating that the Proca field is in the quantum transient
regime according to the results presented in the preceding two sections.

As far as hypothetical particles are concerned, one may consider a massive
photon and use secure upper bounds on its mass discussed in \cite{Nieto_RMP2010}. 
As is reported in Table \ref{ro_tab}, this
leads to  intriguing lower bounds on the oscillation period characterizing 
the quantum transient regime.
These results are striking because they emerge from converting unimaginably
small mass scales into time scales familiar from everyday experience.

\acknowledgments

The research for this publication has been supported by Jagiellonian University.

\appendix

\section{Conventions, continuous-to-discrete transition, and coherent states}
\label{Conventions_app}

{\bf Conventions}.
Unless stated otherwise, the Heaviside-Lorentz system of units is used
in its $\hbar=c=1$ version.  
Greek (Latin) tensor indices 
take values $0,1,2,3$ ($1,2,3$). 
The metric signature is $(+---)$.
$3$-vectors are written in bold, e.g.
$x=(x^\mu)=(x^0,\BD{x})$. The Einstein summation convention is used.
The complex (hermitian) conjugation is denoted by $*$ and $\cc$
($\dag$ and $\hc$). The symbol $\equiv$ denotes a definition,
$\Delta=\bnabla\cdot\bnabla$ represents the Laplacian,
$\Box=\partial_\mu\partial^\mu=\partial_t^2-\Delta$,
$\partial_X=\partial/\partial X$,
 $x^+$ ($x^-$) stands for  a quantity infinitesimally
larger (smaller) than $x$,  and $\BD{\hat{x}}=\BD{x}/|\BD{x}|$.

{\bf The continuous-to-discrete transition}.
The theory can be quantized either in  infinite space or in a periodic box of volume 
$V<\infty$. The transition between these two descriptions is given by the following 
standard mappings
\be 
\int \frac{d^3k}{(2\pi)^{3/2}}a_\KS\leftrightarrow \sum_\BD{k} \frac{b_\KS}{\sqrt{V}},
\ \int d^3k a^\dag_\KS a_\KS \leftrightarrow  \sum_\BD{k} b^\dag_\KS b_\KS, \
\int \frac{d^3k}{(2\pi)^3}
\leftrightarrow  
\sum_{\BD{k}}\frac{1}{V}. 
\ee

{\bf Coherent states}. The following expressions, 
which can be e.g. found in  \cite{Gerry,Scully},  
are needed to   verify  our results
\begin{align}
&D^\dag(\gamma_\KS) b_\KS D(\gamma_\KS) = b_\KS + \gamma_\KS,\\
&\lara{\gamma_\KS|\xi_\KS} =
\exp\B{-\frac{|\gamma_\KS|^2}{2} -   \frac{|\xi_\KS|^2}{2}
+\gamma^*_\KS \xi_\KS},\\
&|\gamma_\KS\ra= \exp\B{- \frac{|\gamma_\KS|^2}{2}} 
\sum_{n=0}^\infty \frac{\gamma_\KS^n}{\sqrt{n!}}|n_\KS\ra,
\label{alfabeta}
\end{align}
where $|n_\KS\ra=(b^\dag_\KS)^n|0_\KS\ra/\sqrt{n!}$ is the $n$-vector boson 
Fock state and $|0_\KS\ra$ is the vacuum state of the  vector boson mode
$\KS$; 
$|0\ra=\bigotimes_\KS|0_\KS\ra$ in the quantization-in-a-periodic-box scheme.

\section{Alternative   derivation of (\ref{h0h1a}) and (\ref{h0h1b}) }
\label{Hellmann_app}

We introduce  the Hamiltonian 
\be
H(\gamma)=H_0+ (1+\gamma) (H_1-\zeta) + (1+\gamma)^2\zeta,
\label{Hgam}
\ee
where $\gamma\in\RR$.
This Hamiltonian  describes the static Proca theory when one
works with the external current $(1+\gamma)\jjmat(\BD{r})$ instead of 
$\jjmat(\BD{r})$; $H(0)$ is equivalent to (\ref{Hdiag}). 
The normalized to  unity 
ground state of (\ref{Hgam}), which we denote as
$|\tilde{\gamma}\ra$, is associated with the  eigenenergy 
\be
\tilde{\cal E}(\gamma)=(1+\gamma)^2\B{
-\sum_\KS\VAREPS{k}|\lambda_\KS|^2 +\zeta
}
\label{Egam}
\ee
that is obtained  by a proper rescaling of (\ref{tildeE}).
The so-called  Hellmann-Feynman theorem guarantees that  
\be
\lara{\tilde{\gamma}|H'(\gamma)|\tilde{\gamma}}=
\tilde{\cal E}'(\gamma),
\ee
where the prime stands for $d/d\gamma$.
Taking  derivatives and  setting $\gamma=0$, we arrive at
\be
\lara{\tilde{0}|H_1-\zeta|\tilde{0}}=-2\sum_\KS\VAREPS{k}|\lambda_\KS|^2, 
\label{jkdfwjhkdfjhdefs}
\ee
which is equivalent to (\ref{h0h1b}).
Equation (\ref{jkdfwjhkdfjhdefs}), when combined with  
$\lara{\tilde{0}|H_0|\tilde{0}} +
\lara{\tilde{0}|H_1-\zeta|\tilde{0}}=-\sum_\KS\VAREPS{k}|\lambda_\KS|^2$
following from the $\gamma=0$ version of (\ref{Egam}), yields a result
equivalent to (\ref{h0h1a}).

\section{Time-dependent  external current}
\label{Time_app}

We provide below  general expressions describing the  Proca field
coupled to a {\it   time-dependent} external current
$\jjmat(t,\BD{r})$.
A similar problem, in the quantum Coulomb-gauge Maxwell theory, was recently treated  in \cite{BDmag}. 
Since it is a straightforward exercise to adapt the formulas from \cite{BDmag} to the present setup,
we will not elaborate on it. 

To proceed, we assume $t\ge0$ in this appendix and provide the expression for 
interaction-picture evolution operator
(\ref{UI})
\begin{align}
\label{UIexplicit}
& U_I(t,0)=\exp\B{-\ii\int_0^t dt' \zeta(t')}\prod_\KS  \exp\B{\ii\text{Im}\Phi_\KS(t)} D\B{\alpha_\KS(t)},\\
& \alpha_\KS(t)= \int_0^t dt' \beta_\KS(t'),\\
& \Phi_\KS(t)=\int_0^t dt'\beta_\KS(t')\int_0^{t'}
dt''\beta_\KS^*(t''),
\end{align}
where 
\begin{align}
\label{betat}
&\beta_\KS(t)=  \frac{-\ii \exp(\ii\VAREPS{k}t) }{\sqrt{2V\VAREPS{k}}}
J_\text{mat}(t,\BD{k})\cdot\eta(\BD{k},\sigma),\\
&J_\text{mat}(t,\BD{k})=\int d^3r \jjmat(t,\BD{r})\exp(-\ii\BD{k}\cdot\BD{r}).
\end{align}
The expression for $\beta_\KS(t)$ matches the one from 
\cite{BDmag} after the replacement of  $\VAREPS{k}$ and
$J_\text{mat}(t,\BD{k})\cdot\eta(\BD{k},\sigma)$ 
by $\OM{k}$ and
$-\BD{J}_\text{mat}(t,\BD{k})\cdot\BD{\eta}(\BD{k},\sigma)$, respectively.
It is evident that it should be so from the comparison of 
\be
H_1^I(t;t)
=\sum_\KS\BB{ \frac{J^*_\text{mat}(t,\BD{k})\cdot\eta(\BD{k},\sigma)}{\sqrt{2V\VAREPS{k}}}
b_\KS \exp(-\ii\VAREPS{k}t)+\hc}+\zeta(t),
\ee
obtained by combining (\ref{H1tt}) and (\ref{A0r}), to the corresponding expression in 
\cite{BDmag}.\footnote{There is no $\zeta(t)$  in \cite{BDmag} but it leads to 
the trivial phase factor in (\ref{UIexplicit}).} 
The discussion of the interaction-picture quantities is 
completed with the following two remarks. First, the
Fock space representation of the interaction-picture 
Proca field is obtained by combining (\ref{V000}) and (\ref{VIII}) 
with (\ref{A0r}). Second, the interaction-picture wavefunction reads 
$|\psi_I(t)\ra = U_I(t,0)|\psi(0)\ra$, which for $|\psi(0)\ra=|0\ra$
reduces to the following  coherent state-based  expression
\be
|\psi_I(t)\ra =
\exp\B{-\ii\int_0^t dt' \zeta(t')}\bigotimes_\KS
\exp\B{\ii\text{Im}\Phi_\KS(t)}  |\alpha_\KS(t)\ra.
\label{cohI}
\ee
Having the above results, one may find the corresponding expressions in other
pictures. In this respect, two results are of interest.

First, using (\ref{V000}), (\ref{VIII}), and  (\ref{A0r}), one may verify that
the Heisenberg-picture Proca field operator, 
$V^0(t,\BD{r};t)= U_I^\dag(t,0) \BD{V}^0_I(t,\BD{r};t) U_I(t,0)$ and 
$\BD{V}(t,\BD{r})= U_I^\dag(t,0) \BD{V}_I(t,\BD{r}) U_I(t,0)$, 
reads
\begin{align}
&V^0(t,\BD{r};t)= 
V^0_\text{free}(t,\BD{r}) +
\sum_{\BD{k}} \frac{ \OM{k}}{\sqrt{2V\VAREPS{k}}m}
\B{\alpha_{\BD{k}3}(t)
\exp(-\ii\VAREPS{k}t+\ii\BD{k}\cdot\BD{r}) 
+\cc}+\frac{\jjmat^0(t,\BD{r})}{m^2},\\
&\BD{V}(t,\BD{r})= 
\BD{V}_\text{free}(t,\BD{r}) +
\sum_\KS \frac{\BD{\eta}(\BD{k},\sigma)   }{\sqrt{2V\VAREPS{k}}}
\B{\alpha_\KS(t)
\exp(-\ii\VAREPS{k}t+\ii\BD{k}\cdot\BD{r}) 
+\cc}.
\end{align}
As a self-consistency check, we have verified that these expressions satisfy 
total current conservation   (\ref{total}).

Second, the Schr\"odinger-picture  wavefunction reads
$|\psi(t)\ra= \exp(-\ii H_0 t)|\psi_I(t)\ra$, 
which  for $|\psi(0)\ra=|0\ra$ results in
\be
|\psi(t)\ra=\exp\B{-\ii\int_0^t dt' \zeta(t')}\bigotimes_\KS \exp\B{\ii\text{Im}\Phi_\KS(t)} 
|\alpha_\KS(t)\exp(-\ii\VAREPS{k}t) \ra.
\label{SchSch}
\ee

Finally, we note that the above findings generalize the
textbook results on quantum dynamics induced by a time-dependent external
current in a massless theory \cite{Gerry,Scully,PandA}. To the best of our
knowledge, such a generalization has not been reported previously. We further
emphasize that the massless counterpart of the phase factor associated with
$\Phi_\KS$ is omitted  in \cite{Gerry,Scully,PandA}.

\section{Riemann–Lebesgue-like  lemma}
\label{RL_app}

We consider an integral 
\be
I(t)=\int_0^\infty d\OM{k} f(\OM{k}) \cos(\VAREPS{k}t),
\label{w1elfwle}
\ee
where $f$ is real-valued. 
After the change of the integration variable, we obtain 
\be
I(t)=\int_{m}^\infty dx F(x) \cos(x t),
\ F(x)=\frac{x}{\sqrt{x^2-m^2}} f\B{\sqrt{x^2-m^2}}.
\ee
Then, according to the Riemann-Lebesgue lemma, we have
\be
\lim_{t\to\infty} I(t)= 0 
\ee
for real-valued $F\in L^1(m,\infty)$. This is  equivalent to requiring that 
$f\in L^1(0,\infty)$, which for $f$ that  does not change sign 
corresponds to $|\int_0^\infty d\OM{k} f(\OM{k})|<\infty$. 
Since the very  same reasoning can be 
repeated after the replacement of 
$ \cos(\VAREPS{k}t)$ by $\sin(\VAREPS{k}t)$ in (\ref{w1elfwle}),
we conclude that (\ref{RLL}) holds.

Finally, we note that when we invoke (\ref{RLL}) in the main body of the paper,
we deal with expressions given by a sum over $\BD{k}$. This is done under the 
implicit assumption that before using (\ref{RLL}), one replaces this
sum  by the integral and  performs the angular integration
in momentum space.

\section{Derivation of expanding point-charge solution}
\label{Integrals_app}

We combine (\ref{v000}) with
(\ref{lks}) and (\ref{V0stat}) getting
\be
\lara{\psi(t)|V^0(0,\BD{r};t)|\psi(t)}=\frac{q}{4\pi r}\exp(-m r)+I_1(t,\BD{r}) + I_2(t,\BD{r}),
\label{eqq1}
\ee
where
\begin{align}
&I_1(t,\BD{r})=-q \int \frac{d^3k}{(2\pi)^3}
\frac{\exp(-\ii\VAREPS{k}t+\ii\BD{k}\cdot\BD{r})}{2\VAREPS{k}^2}+\cc,\\
&I_2(t,\BD{r})=\frac{q}{m^2}\int  \frac{d^3k}{(2\pi)^3}
\frac{\exp(-\ii\VAREPS{k}t+\ii\BD{k}\cdot\BD{r})}{2}+\cc
\end{align}
Regarding  the former integral, we note that $\partial_t
I_1(t,\BD{r})=-q\Delta(t,\BD{r})$. Thereby, 
\be
I_1(t,\BD{r})=-\frac{q}{4\pi r}\exp(-m r)-q\int_0^t dt' \Delta(t',\BD{r}),
\label{ed1}
\ee
where the first term represents   $I_1(0,\BD{r})$.
Regarding the latter integral, we note that
\be
I_2(t,\BD{r})=-\frac{q}{m^2} \partial_t\Delta(t,\BD{r}).
\label{ed2}
\ee
Putting  (\ref{ed1}) and (\ref{ed2})  into (\ref{eqq1}), we obtain  (\ref{V0tr}).

Then, we combine 
 (\ref{vvvv}) with  (\ref{lks}) and   (\ref{Vstat}) getting 
\be
\lara{\psi(t)|\BD{V}(0,\BD{r})|\psi(t)} =
\frac{q}{2m^2}\int \frac{d^3k}{(2\pi)^3}\frac{\BD{k}}{\VAREPS{k}}
\exp(-\ii\VAREPS{k}t+\ii\BD{k}\cdot\BD{r})+\cc
\ee
It follows from the definition of
$\Delta(t,\BD{r})$ that this result  is equivalent  to (\ref{Vvtr}).

\section{Self-consistency check of expanding point-charge solution}
\label{Solution_app}

It follows from field equations that
the  Heisenberg-picture identity,
$ \partial\cdot V=\partial\cdot\jjmat/m^2$,
must    hold.
Taking the  expectation value
of this  expression in the
Heisenberg-picture state,
and then rewriting the resulting expression to 
the Schr\"odinger picture, we arrive at  
\be
\label{r3finjrf}
\partial_t \lara{\psi(t)|V^0(0,\BD{r};t)|\psi(t)}+
\bnabla\cdot\lara{\psi(t)|\BD{V}(0,\BD{r})|\psi(t)}=\frac{\partial\cdot\jjmat(t,\BD{r})}{m^2}.
\ee
To verify it, we note that
for any $t$, not just $t>0$ as in (\ref{V0tr}) and (\ref{Vvtr}), we have 
\begin{align}
&\lara{\psi(t)|V^0(0,\BD{r};t)|\psi(t)}=\theta(t) \BB{-q\int_0^t dt'\Delta(t',\BD{r})
-\frac{q}{m^2}\partial_t\Delta(t,\BD{r})},\\
&\lara{\psi(t)|\BD{V}(0,\BD{r})|\psi(t)} = 
\theta(t)\frac{q}{m^2}\bnabla\Delta(t,\BD{r}).
\end{align}
Using these expressions,
we find that the left-hand side of (\ref{r3finjrf})
reads 
\be
\frac{q\delta(\BD{r})\delta(t)}{m^2}-\theta(t)\frac{q}{m^2}\B{\Box+m^2}\Delta(t,\BD{r}),
\ee
which  agrees with the right-hand side of (\ref{r3finjrf}) for
$\jjmat(t,\BD{r})$ given by (\ref{JmatMAGt}) because 
$\Delta(t,\BD{r})$ satisfies the Klein-Gordon equation.
Finally, we note that the above discussion is formal due to 
the divergence of the underlying expressions at   $r=t$.

\section{Derivation of (\ref{InsA})}
\label{Derivation_app}

We assume $r<t$ and note that it follows from (\ref{PJ}) that  
\be
-q\int_0^t dt' \Delta(t',\BD{r})=\frac{q}{4\pi r} + 
\frac{qm}{4\pi}\int_{mt}^\infty dx \frac{\BesselJ_1\B{\sqrt{x^2-(mr)^2}}}{\sqrt{x^2-(mr)^2}} + I,
\label{wemll}
\ee
where 
\be
I=-\frac{qm}{4\pi}\int_{mr}^\infty dx
\frac{\BesselJ_1\B{\sqrt{x^2-(mr)^2}}}{\sqrt{x^2-(mr)^2}}.
\ee
Using  standard 
 Bessel function identities, 
 we
rewrite this expression to the following form
\be
I=-\frac{q}{4\pi r} \B{1+\frac{d}{d(mr)}\int_0^\infty dz \frac{z
\BesselJ_0(z)}{\sqrt{z^2+(mr)^2}}},
\label{edlk}
\ee
where the integral  equals $\exp(-mr)$
according 		to
formula 6.554.1 of \cite{Ryzhik}. Putting this  result into (\ref{edlk}), 
and then the resulting expression into (\ref{wemll}), 
one obtains  (\ref{InsA}).

\section{Derivation of (\ref{Posc})}
\label{Der_app}
We set $mr$  to zero 
in the last three terms  of (\ref{Ins}).
Their sum then reads  
\be
\frac{qm}{4\pi}I-\frac{q}{4\pi t}\BesselJ_0(mt)
+\frac{q}{2\pi m t^2}\BesselJ_1(mt),
\label{weffre}
\ee
where
\be
I=\int_{mt}^\infty dx \frac{\BesselJ_1(x)}{x}. 
\ee
Using  standard Bessel function identities, we arrive at
\be
I=\int_0^\infty dx \BesselJ_0(x)
-\int_0^{mt} dx \BesselJ_0(x)+
\BesselJ_1(mt),
\ee
where the first (second) integral is equal to $1$ 
($mt\BesselJ_0(mt)+\pi mt\BesselJ_1(mt)\StruveH_0(mt)/2-
\pi mt\BesselJ_0(mt)\StruveH_1(mt)/2$)
according to formula 6.511.1 (6.511.6)  of
\cite{Ryzhik}; $\StruveH_n$ 
represents the Struve function of order $n$.
Substituting the resulting expression  for $I$ into (\ref{weffre}),
and using the standard asymptotic expressions for 
$\BesselJ_n$ and $\StruveH_n$  \cite{Ryzhik}, 
the right-hand side of (\ref{Posc}) is obtained. 

\section{Exact results   for expanding extended-charge problem}
\label{IntegralSTAT_E_app}

{\bf The energy}. Using (\ref{Hall}), (\ref{zetA}), 
(\ref{jrs}), and (\ref{lkewfkw}), we arrive at  
\be
{\cal E}= \frac{q^2}{(2\pi m\epsilon)^2\epsilon}\int_0^\infty dx x^2
\exp\B{-\frac{x^2}{2}},
\ee
which  is evaluated via formula 3.461.2b of \cite{Ryzhik}
\be
\int_0^\infty dx x^m\exp\B{-\frac{x^2}{2}}=2^{(m-1)/2}
\Gamma\B{\frac{m+1}{2}},
\label{INTAPP}
\ee
where $\Gamma$ is  the gamma function.
This leads to (\ref{Estt}).

{\bf The standard deviation of the energy}. Using (\ref{sigE}),
(\ref{hnn}),  and (\ref{lksEX}), we arrive at 
\begin{align}
&\sigma^2_{\cal E}=\frac{q^2}{(2\pi \ME)^2 \epsilon^2}I,\\
&I= \int_0^\infty dx \frac{x^4\exp(-x^2/2)  }{\sqrt{(\ME)^2+ x^2}}.
\label{2I2}
\end{align}
To evaluate  $I$, we observe that 
\be
I=\left.\frac{d^2}{dp^2}\int_0^\infty dx \frac{\exp(-p x^2)  }{\sqrt{(\ME)^2+ x^2}}\right|_{p=1/2}
\label{2I2_a}
\ee
and then note that according to formula 3.462.25 of \cite{Ryzhik},
the integral in the above equation equals 
\be
\frac{1}{2}\exp\B{\frac{p(\ME)^2}{2}}\BesselK_0\B{\frac{p(\ME)^2}{2} }.
\label{2I2_b}
\ee
Using standard Bessel function identities, we 
have obtained (\ref{dew}) from the above  expressions.

\section{Exact results for static extended-charge problem}
\label{IntegralSTAT_S_app}

{\bf The expectation value of $V^0$}. Using  (\ref{Vstat0}) 
and (\ref{jrs})--(\ref{lksEX}),
we arrive at 
\begin{align}
&\lara{\tilde{\psi}(t)|V^0(0,\BD{r})|\tilde{\psi}(t)} =
\frac{q\exp(-r^2/\epsilon^2)}{\pi^{3/2}(m\epsilon)^2\epsilon}
- \frac{q}{2\pi^2 ( m\epsilon)^2 r} I_0, \\
&I_0=\int_0^\infty dx \frac{x^3\exp(-x^2/4) }{(\ME)^2+ x^2}
\sin\B{x \frac{r}{\epsilon}}.
\end{align}
To evaluate $I_0$, we observe  that 
\begin{subequations}
\begin{align}
\label{I0a}
I_0&=\int_0^\infty dx x \exp(-x^2/4)\sin\B{x \frac{r}{\epsilon}}\\
&-(\ME)^2 \int_0^\infty dx
\frac{x\exp(-x^2/4)  }{(\ME)^2+x^2}
\sin\B{x \frac{r}{\epsilon}}.
\label{I0b}
\end{align}
\end{subequations}
Expressions  (\ref{I0a}) and (\ref{I0b}) are evaluated via
formulas 3.952.1 and 3.954.1  of \cite{Ryzhik}, respectively.
The former expression equals 
\be
2\sqrt{\pi}\frac{r}{\epsilon}\exp\B{-\frac{r^2}{\epsilon^2}}
\ee
while the latter 
\be
\frac{\pi}{4}(\ME)^2\exp\B{\frac{(\ME)^2}{4}}\BB{\exp(m r)\erfc\B{\frac{\ME}{2}+\frac{r}{\epsilon}}
-\exp(-m r)\erfc\B{\frac{\ME}{2}-\frac{r}{\epsilon}}}.
\ee
Using these results, we have obtained (\ref{efw}).

{\bf The ground-state energy}. Using (\ref{tildeE}), (\ref{lksEX}),  (\ref{Hall}), and 
(\ref{Estt}), we  arrive at the following expression 
\begin{align}
&\tilde{\cal E} =
\frac{q^2}{(2\pi \ME)^2\epsilon}\BB{\sqrt{\frac{\pi}{2}}
-I_1},\\
&I_1=\int_0^\infty dx \frac{x^4\exp(-x^2/2)  }{(\ME)^2+ x^2}.
\end{align}
To evaluate  $I_1$, we observe that 
\begin{align}
I_1&=\int_0^\infty dx
\B{x^2-(\ME)^2+ \frac{(\ME)^4}{(\ME)^2+x^2}}\exp\B{-\frac{x^2}{2}}  \nonumber\\
& =\sqrt{\frac{\pi}{2}}-(\ME)^2\sqrt{\frac{\pi}{2}} +
(\ME)^4\int_0^\infty dx \frac{\exp(-x^2/2) }{(m\epsilon)^2+ x^2}, 
\label{dekllkq}
\end{align}
where we have used (\ref{INTAPP}).  The remaining integral reads 
\be
\frac{\pi}{2\ME} \exp\B{\frac{(\ME)^2}{2}} \erfc\B{\frac{\ME}{\sqrt{2}}}
\ee
according to formula 3.466.1 of \cite{Ryzhik}. These findings lead to 
(\ref{wwds}).

{\bf The vector-boson  number}. Using
(\ref{phn}) and  (\ref{lksEX}), 
we arrive at 
\begin{align}
&\tilde{\cal  N}=\frac{q^2}{(2\pi \ME)^2}I_2,\\
&I_2= \int_0^\infty dx \frac{x^4\exp(-x^2/2)  }{\BB{(\ME)^2+ x^2}^{3/2}}.
\end{align}
To evaluate $I_2$, we observe that 
\be
I_2=-\frac{1}{\ME}\frac{d}{d(\ME)} I,
\ee
where $I$ is given by (\ref{2I2}). Then, 
we use   (\ref{2I2_a}) and (\ref{2I2_b}) to obtain  (\ref{ewuy})
via standard Bessel function identities.



%

\end{document}